\documentclass[apj]{emulateapj}

\usepackage{multirow}
\usepackage{booktabs}
\usepackage{array}

\usepackage[colorlinks=true,citecolor=blue]{hyperref}

\DeclareMathAlphabet{\mathsc}{OT1}{cmr}{m}{sc}
\def\testbx{bx}
\DeclareRobustCommand{\ion}[2]{
\relax\ifmmode
\ifx\testbx\f@series
{\mathbf{#1\,\mathsc{#2}}}\else
{\mathrm{#1\,\mathsc{#2}}}\fi
\else\textup{#1\,{\mdseries\textsc{#2}}}
\fi}
\newcommand{\Hi} {\ion{H}{i}}

\newcommand{\ha} {\mbox{H$\alpha$}}
\newcommand{\hb} {\mbox{H$\beta$}}
\newcommand{\hg} {\mbox{H$\gamma$}}

\newcommand{\Hei} {\ion{He}{i}}

\newcommand{\Oia} {[\ion{O}{i}]}

\newcommand{\Nai} {\ion{Na}{i}}

\newcommand{\Siii} {\ion{Si}{ii}}

\newcommand{\Caiia} {[\ion{Ca}{ii}]}
\newcommand{\Caii} {\ion{Ca}{ii}}
\newcommand{\Scii} {\ion{Sc}{ii}}
\newcommand{\Tiii} {\ion{Ti}{ii}}

\newcommand{\Feii} {\ion{Fe}{ii}}

\newcommand{\Baii} {\ion{Ba}{ii}}

\newcommand{\sn}{SN 2013ej}
\newcommand{\host}{NGC 0628}
\newcommand{\EpEpoch}{JD 2456497.30}

\newcommand{\synow}{\textsc{synow}}

\newcommand{\iraf}{\texttt{IRAF}}

\newcommand{\cmfgen}{\textsc{cmfgen}}

\newcommand{\epm}{\textsc{EPM}}

\newcommand{\ebv}{\mbox{$E(B-V)$}}

\newcommand{\maghundred}{\mbox{mag (100 d)$ ^{-1} $}}
\newcommand{\msun}{\mbox{M$_{\odot}$}}

\newcommand{\rsun}{\mbox{R$_{\odot}$}}
\newcommand{\kms}{\mbox{$\rm{\,km\,s^{-1}}$}}
\newcommand{\ergs}{\mbox{$\rm{\,erg\,s^{-1}}$}}

\newcommand{\nickel}{\mbox{$^{56}$Ni}}
\newcommand{\cobalt}{\mbox{$^{56}$Co}}
\newcommand{\iron}{\mbox{$^{56}$Fe}}
\newcommand{\mum}{\mbox{$\mu{\rm m}$}}

\newcommand{\ld}{\mbox{$\lambda$}}
\newcommand{\ldld}{\mbox{$\lambda\lambda$}}

\shorttitle{Type IIL Supernova 2013ej}
\shortauthors{Bose et al.}

\begin{document}

\title{\sn\ - A type IIL supernova with weak signs of interaction}

\author{
Subhash Bose$^{\star}$\altaffilmark{1,2},
Firoza Sutaria\altaffilmark{3},
Brijesh Kumar\altaffilmark{1},
Chetna Duggal\altaffilmark{3},
Kuntal Misra\altaffilmark{1},
Peter J. Brown\altaffilmark{4},
Mridweeka Singh\altaffilmark{1},
Vikram Dwarkadas\altaffilmark{5},
Donald G. York\altaffilmark{6},
Sayan Chakraborti\altaffilmark{7},
H.C. Chandola\altaffilmark{2},
Julie Dahlstrom\altaffilmark{8},
Alak Ray\altaffilmark{9} and
Margarita Safonova\altaffilmark{3}
}

\email{$^\star$email@subhashbose.com, bose@aries.res.in}

\altaffiltext{1}{Aryabhatta Research Institute of Observational Sciences, Manora
    Peak, Nainital 263002, India}
\altaffiltext{2}{Centre of Advance Study, Department of Physics, Kumaun University, Nainital - 263001, India.}
\altaffiltext{3}{Indian Institute of Astrophysics, Block-II, Koramangala, Bangalore - 560034, India.}
\altaffiltext{4}{George P. and Cynthia Woods Mitchell Institute for Fundamental
Physics \& Astronomy, Texas A. \& M. University, Department
of Physics and Astronomy, 4242 TAMU, College Station,
TX 77843, USA}
\altaffiltext{5}{Department of Astronomy and Astrophysics, University of Chicago, Chicago, Illinois, 60637, USA}
\altaffiltext{6}{Department of Astronomy and Astrophysics and The Enrico Fermi Institute, University of Chicago, 60637, USA}
\altaffiltext{7}{Institute for Theory and Computation, Harvard–Smithsonian Center for Astrophysics, 60 Garden Street, Cambridge, MA 02138, USA}
\altaffiltext{8}{Carthage College, 2001 Alford Park Dr., Kenosha, WI 53140, USA}
\altaffiltext{9}{Tata Institute of Fundamental Research, Homi Bhabha Road, Mumbai 400005}

\begin{abstract}
We present optical photometric and spectroscopic observations of supernova 2013ej. It is one of the brightest type II supernovae exploded in a nearby ($\sim 10$ Mpc) galaxy NGC 628. The light curve characteristics are similar to type II SNe, but with a relatively shorter ($ \sim85 $ day) and steeper ($ \sim1.7 $ mag (100 d)$^{-1} $ in \textit{V}) plateau phase. The SN shows a large drop of 2.4 mag in \textit{V} band brightness during plateau to nebular transition. The absolute ultraviolet (UV) light curves are identical to SN 2012aw, showing a similar UV plateau trend extending up to 85 days. The radioactive \nickel\ mass estimated from the tail luminosity is $ 0.02 $\msun\ which is significantly lower than typical type IIP SNe. The characteristics of spectral features and evolution of line velocities indicate that SN 2013ej is a type II event. However, light curve characteristics and some spectroscopic features provide strong support in classifying it as a type IIL event. A detailed \synow\ modelling of spectra indicates the presence of some high velocity components in \ha\ and \hb\ profiles, implying possible ejecta-CSM interaction. The nebular phase spectrum shows an unusual notch in the \ha\ emission which may indicate bipolar distribution of \nickel. Modelling of the bolometric light curve yields a progenitor mass of  $ \sim14 $\msun\ and a radius of $ \sim450 $\rsun, with a total explosion energy of $ \sim2.3\times10^{51} $ erg.

\end{abstract}


\section{Introduction} \label{sec:intro}

Type II supernovae (SNe) originate from massive stars with $ M_{ZAMS} > 8 \msun $ \citep{2013RvMP...85..245B} which have retained substantial hydrogen in the envelope at the time of explosion. They belong to a subclass of core-collapse SNe (CCSNe), which collapse under their own gravity at the end of the nuclear burning phase, having insufficient thermal energy to withstand the collapse.

The most common subtype among hydrogen rich supernovae is type IIP. At the time of shock breakout almost the entire mass of hydrogen is ionized. Type IIP SNe have an extended hydrogen envelope, which recombines slowly over a prolonged duration sustaining the plateau phase. During this phase the SN light curve shows almost constant brightness lasting for 80-100 days. At the end of plateau phase the SN experiences a sudden drop in luminosity, settling onto the slow declining radioactive tail, also known as nebular phase, which is mainly powered by gamma rays released from the decay of \cobalt\ to \iron, which in turn depends upon the amount of \nickel\ synthesized at the time of explosion.

{The plateau slope of SN type II light curve primarily depends on the amount of hydrogen present in the ejecta. If hydrogen content is high, as in type IIP, the initial energy deposited from shock and decay of freshly produced \nickel\, shall be released slowly over a longer period of time. On the other hand if hydrogen content is relatively low, the light curve will decline fast but with higher peak luminosity. Thus if hydrogen content is low enough, one would expect a linear decline in the light curve classifying it as type IIL.
By the historical classification, type IIL \citep{1979A&A....72..287B} shows linear decline in light curve over 100 days until it reaches the radioactive tail phase. \cite{2012ApJ...756L..30A} claimed to find type IIP and IIL as to distinct group of events which may further indicate their distinct class of progenitors. However, recent studies by \cite{2014ApJ...786...67A} and \cite{2015ApJ...799..208S} on large sample of type II SNe do not favor any such bi-modality in the diversity, rather they found continuum in light curve slopes as well as in other physical parameters. The continuous distribution of plateau slopes in type II events is rather governed by variable amount of hydrogen mass left in the envelope at the time of explosion. Based on a sample of 11 type IIL events, \cite{2014MNRAS.445..554F} proposed that any event having decline of 0.5mag in V band light curve in first 50 days can be classified as type IIL. In light of these recent developments a large number of type IIP SNe classified earlier may now fall under IIL class. Thus many of the past studies collectively on samples of type IIP SNe, which we shall be referring in this work may include both IIP as well as IIL.}

Extensive studies have been done to relate observable parameters and progenitor properties of IIP SNe \citep[e.g.,][]{1985SvAL...11..145L,2003ApJ...582..905H}. Stellar evolutionary models suggest that these SNe may originate from stars with zero-age-main-sequence mass of 9-25\msun\ \citep[e.g.,][]{2003ApJ...591..288H}. However, progenitors directly recovered for a number of nearby IIP SNe, using the pre-SN \textit{HST} archival images, are found to lie within $ 8-17 \msun$ RSG stars \citep{2009ARA&A..47...63S}. Recent X-ray study also infers an upper mass limit of $ <19\msun $ for type IIP progenitors \citep{2014MNRAS.440.1917D}, which is in close agreement to that obtained from direct detection of progenitors.

The geometry of the explosion and presence of pre-existent circumstellar medium (CSM), often associated with progenitor mass loss during late stellar evolutionary phase, can significantly alter the observables even though originating from similar progenitors. There are number of recent studies of II SNe, like 2007od \citep{2011MNRAS.417..261I}, 2009bw \citep{2012MNRAS.422.1122I} and 2013by \citep{2015arXiv150106491V} which show signature of such CSM interactions during various phases of evolution.

\sn\ is one of the youngest detected type II SN which was discovered soon after its explosion. The earliest detection was reported on July 24.125 UTC, 2013 by C. Feliciano in \textit{Bright Supernovae\footnote{http://www.rochesterastronomy.org/supernova.html}} and subsequent independent detection on July 24.83 UTC by \cite{2013ATel.5466....1L} at \textit{V}-band magnitude of $ \sim $14.0. The last non-detection was reported on July 23.54 UTC, 2013 by All Sky Automated Survey
for Supernovae \citep{2013ATel.5237....1S} at a \textit{V}-band detection limit of $ > 16.7 $ mag. Therefore, we adopt an explosion epoch (0d) of July 23.8 UTC (JD $ =2456497.3\pm0.3 $), which is chosen in between the last non-detection and first detection of \sn. This being one of the nearest and brightest events, it provides us with an excellent opportunity to study the origin and evolution of type II SN. Some of the basic properties of \sn\ and its host galaxy are listed in Table~\ref{tab:host}.

\cite{2014MNRAS.438L.101V} presented early observations of \sn\ and using temperature evolution for the first week, they estimated a progenitor radius of 400-600 \rsun. \cite{2014MNRAS.439L..56F} used high resolution archival images from \textit{HST} to examine the location of \sn\ and identified the progenitor candidate to be a supergiant of mass $ 8-15.5\msun $. \cite{2013ATel.5275....1L} reported unusually high polarization using spectropolarimetric observation for the week old SN, as implying substantial asymmetry in the scattering atmosphere of ejecta. {X-ray emission has also been detected by \textit{Swift} XRT \citep{2013ATel.5243....1M}, which may indicate \sn\ has experienced CSM interaction.}

In this work we present photometric and spectroscopic observation of \sn, and carry out qualitative as well as quantitative analysis of the various observables through modelling and comparison with other archetypal SNe. The paper is organized as follows. In section \ref{sec:obs} we describe photometric and spectroscopic observations and data reduction. The estimation of line of sight extinction is discussed in section \ref{sec:ext}. In section \ref{sec:lc} we analyze the light curves, compare absolute magnitude light curves and color curves. We also derive bolometric luminosities and estimate nickel mass from the tail luminosity. Optical spectra are analyzed in  section \ref{sec:sp}, where we model and discuss evolution of various spectral features and compare velocity profile with other type II SNe. In section \ref{modelling}, we model the bolometric light curve of \sn\ and estimate progenitor and explosion parameters. Finally in section \ref{sec:sum}, we summarize the results of this work.


  \begin{table}
  \caption{Relevant parameters for the host galaxy \host\ and \sn.}
  \label{tab:host} 

  \begin{tabular}{llc} \hline \hline
     \noalign{\smallskip}
      Parameters& Value& Ref.$^{a}$\\ 
     \noalign{\smallskip} \hline
    
     \noalign{\smallskip}
     \multicolumn{3}{l}{\bf \host:}\\
     Alternate name& M74& 2\\
     Type& Sc& 2\\
     RA (J2000)& $\alpha = 01^{\rm h} 36^{\rm m} 41\fs77$& 2\\
     DEC (J2000)& $\delta = 15\degr 46\arcmin 59\farcs8$& 2\\
     Abs. Magnitude& $M_{B}=-20.72$ mag& 2\\
     \\
     Distance& $D=9.6\pm0.7$ Mpc& 1 \\
     Distance modulus& $\mu = 29.90\pm0.16$ mag& \\
     \\
     Heliocentric Velocity& $cz_{\rm helio}=658\pm1 \kms$&2\\
     \\
     \multicolumn{3}{l}{\bf \sn:}\\
     RA (J2000)& $\alpha = 01^{\rm h} 36^{\rm m} 48\fs16$& 3\\
     DEC (J2000)& $\delta = 15\degr 45\arcmin 31\farcs3$& \\
     \\
     Galactocentric Location& 1\arcmin33\arcsec E, 2\arcmin15\arcsec S&  \\
     \\
     Date of explosion & $t_{\rm 0}$ =23.8 July 2013 (UT)& 1\\ 
                    & (JD $ 2456497.3\pm0.3 $)& \\ 
     Reddening & \ebv\,$=0.060\pm0.001$ mag& 1\\
   
     \noalign{\smallskip}
     \hline
  \end{tabular}
  \newline (1) This paper; 
           (2) HyperLEDA - http://leda.univ-lyon1.fr;
           (3) \citet{2013CBET.3606....1K}
  \end{table}

\section{Observation and data reduction} \label{sec:obs}

\subsection{Photometry} \label{sec:obs.phot}
Broadband photometric observations in \textit{UBVRI} filters have been carried out from 2.0m IIA Himalayan Chandra Telescope (HCT) telescope at Hanle and ARIES 1.0m Sampurananand (ST) and 1.3m Devasthal Fast Optical (DFOT) telescopes at Nainital. Additionally \sn\ has been also observed with \textit{Swift} Ultraviolet/optical (UVOT) telescope in all six bands.

Photometric data reductions follows the same procedure as described in \cite{2013MNRAS.433.1871B}. Images are cleaned and processed using standard procedures of IRAF software. DAOPHOT routines have been used to perform PSF photometry and extracting differential light-curves.
To standardize the SN field, three Landolt standard fields (PG~0231, PG~2231 and SA~92) were observed on October 27, 2013 with 1.0-m ST under good photometric night and seeing (typical FWHM $ \sim$2\arcsec.1 in \textit{V} band) condition. For atmospheric extinction measurement, PG~2231 and PG~0231 were observed at different air masses. The SN field has been also observed in between standard observations. The standardization coefficients derived are represented in the following transformation equations,

\begin{figure}
\centering
\includegraphics[width=8.4cm]{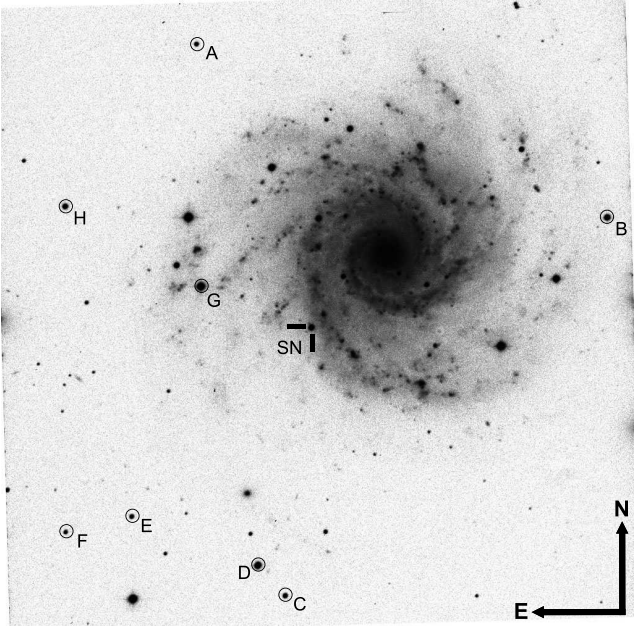}
\caption{\sn\ in \host. The $BR$-band composite image taken from 104-cm Sampurnanad telescope, covering an area of
         about 13\arcmin$\times$13\arcmin\, is shown. Eight local field standards and SN are marked in the image.}
\label{fig:snid}
\end{figure}

\begin{table*}
 \centering
  \caption{Eight local standards in the field of \sn\ with corresponding coordinates ($\alpha, \delta$) calibrated magnitudes in \textit{UBVRI}.  Errors quoted here, include both photometric and calibration errors.}
  \label{tab:photstar}

  \begin{tabular}{cccccccc}
     \hline
     Star& $\alpha_{\rm J2000}$&       $\delta_{\rm J2000}$& $U$& $B$&  $V$&  $R$&  $I$\\
       ID&          (h m s)&(\degr\,\arcmin\,\arcsec)&(mag)&(mag)&(mag)&(mag)&(mag)\\
     \hline
     A& 1:36:57.9 & +15:51:19.4 & 16.773 $\pm$ 0.0325 & 16.867 $\pm$ 0.0259 & 16.297 $\pm$ 0.0193 & 15.939 $\pm$ 0.0163  & 15.567 $\pm$ 0.0207 \\
     B& 1:36:23.0 & +15:47:45.3 & 15.102 $\pm$ 0.0289 & 15.109 $\pm$ 0.0302 & 14.580 $\pm$ 0.0234 & 14.253 $\pm$ 0.0183  & 13.888 $\pm$ 0.0260 \\
     C& 1:36:50.4 & +15:40:01.9 & 16.588 $\pm$ 0.0318 & 16.525 $\pm$ 0.0277 & 15.798 $\pm$ 0.0200 & 15.372 $\pm$ 0.0157  & 14.925 $\pm$ 0.0187 \\
     D& 1:36:52.7 & +15:40:39.4 & 14.811 $\pm$ 0.0318 & 14.561 $\pm$ 0.0265 & 13.817 $\pm$ 0.0184 & 13.384 $\pm$ 0.0167  & 12.976 $\pm$ 0.0196 \\
     E& 1:37:03.4 & +15:41:39.2 & 17.140 $\pm$ 0.0386 & 17.064 $\pm$ 0.0251 & 16.407 $\pm$ 0.0200 & 16.008 $\pm$ 0.0160  & 15.601 $\pm$ 0.0206 \\
     F& 1:37:09.0 & +15:41:20.4 & 18.804 $\pm$ 0.1537 & 17.800 $\pm$ 0.0282 & 16.769 $\pm$ 0.0249 & 16.146 $\pm$ 0.0167  & 15.583 $\pm$ 0.0205 \\
     G& 1:36:57.6 & +15:46:22.7 & 13.934 $\pm$ 0.0272 & 13.756 $\pm$ 0.0219 & 12.991 $\pm$ 0.0160 & 12.555 $\pm$ 0.0161  & 12.155 $\pm$ 0.0240 \\
     H& 1:37:09.0 & +15:48:00.6 & 16.974 $\pm$ 0.0434 & 16.172 $\pm$ 0.0249 & 15.175 $\pm$ 0.0157 & 14.598 $\pm$ 0.0170  & 14.062 $\pm$ 0.0194 \\
     \hline
  \end{tabular}

\end{table*}

\begin{table*}
\centering
  \fontsize{2.0mm}{2.6mm}\selectfont
  \caption{Photometric evolution of \sn. Errors denote $1\sigma$ uncertainty.}
  \label{tab:photsn}
  \textit{UBVRI} photometry\\
  \begin{tabular}
  {c c r c c c c c l c}
  \hline
  UT Date&JD&Phase$^{a}$&$U$&$B$&$V$&$R$&$I$&Tel$^{b}$ \\
  (yyyy-mm-dd)&2456000+&(day)&(mag)&(mag)&(mag)&(mag)&(mag)& \\
  \hline

2013-08-04.82  &   509.32  &    12.02  &  12.026 $\pm$ 0.061  &  12.633 $\pm$ 0.020  &  12.612 $\pm$ 0.013  &  12.434 $\pm$ 0.017  &  12.349 $\pm$ 0.018  &  HCT \\
2013-08-31.93  &   536.43  &    39.13  &  14.576 $\pm$ 0.251  &  14.208 $\pm$ 0.020  &  13.125 $\pm$ 0.011  &  12.670 $\pm$ 0.015  &  12.436 $\pm$ 0.011  &  HCT \\
2013-09-29.77  &   565.27  &    67.97  &  16.088 $\pm$ 0.027  &  14.991 $\pm$ 0.020  &  13.569 $\pm$ 0.012  &  13.056 $\pm$ 0.016  &  12.750 $\pm$ 0.016  &  HCT \\
2013-09-30.72  &   566.22  &    68.92  &  16.207 $\pm$ 0.109  &  14.956 $\pm$ 0.020  &  13.595 $\pm$ 0.008  &  12.992 $\pm$ 0.015  &  12.741 $\pm$ 0.016  &  ST \\
2013-10-02.87  &   568.37  &    71.07  &  16.223 $\pm$ 0.028  &  15.017 $\pm$ 0.020  &  13.640 $\pm$ 0.012  &  13.053 $\pm$ 0.015  &  12.750 $\pm$ 0.017  &  HCT \\
2013-10-13.70  &   579.20  &    81.90  &  16.823 $\pm$ 0.054  &  15.291 $\pm$ 0.015  &  13.864 $\pm$ 0.010  &  13.222 $\pm$ 0.017  &         ---          &  ST \\
2013-10-15.85  &   581.35  &    84.05  &  17.026 $\pm$ 0.061  &  15.365 $\pm$ 0.021  &  13.884 $\pm$ 0.012  &  13.273 $\pm$ 0.011  &  12.978 $\pm$ 0.017  &  ST \\
2013-10-16.71  &   582.21  &    84.91  &  17.036 $\pm$ 0.089  &  15.406 $\pm$ 0.013  &  13.939 $\pm$ 0.010  &  13.288 $\pm$ 0.018  &  12.986 $\pm$ 0.025  &  ST \\
2013-10-21.73  &   587.23  &    89.93  &  17.292 $\pm$ 0.057  &  15.611 $\pm$ 0.017  &  14.126 $\pm$ 0.017  &  13.446 $\pm$ 0.016  &  13.147 $\pm$ 0.018  &  ST \\
2013-10-24.70  &   590.20  &    92.90  &  17.405 $\pm$ 0.035  &  15.743 $\pm$ 0.016  &  14.233 $\pm$ 0.021  &  13.540 $\pm$ 0.014  &  13.253 $\pm$ 0.014  &  ST \\
2013-10-25.72  &   591.22  &    93.92  &  17.365 $\pm$ 0.023  &  15.732 $\pm$ 0.014  &  14.340 $\pm$ 0.008  &  13.592 $\pm$ 0.011  &  13.322 $\pm$ 0.011  &  DFOT \\
2013-10-26.74  &   592.24  &    94.94  &  17.442 $\pm$ 0.020  &  15.795 $\pm$ 0.014  &  14.431 $\pm$ 0.007  &  13.672 $\pm$ 0.010  &  13.384 $\pm$ 0.011  &  DFOT \\
2013-10-27.76  &   593.26  &    95.96  &  17.515 $\pm$ 0.033  &  15.985 $\pm$ 0.022  &  14.453 $\pm$ 0.016  &  13.750 $\pm$ 0.016  &  13.447 $\pm$ 0.021  &  ST \\
2013-11-09.63  &   606.13  &   108.83  &  18.440 $\pm$ 0.039  &  17.611 $\pm$ 0.015  &  16.108 $\pm$ 0.012  &  15.144 $\pm$ 0.015  &  14.783 $\pm$ 0.016  &  HCT \\
2013-11-11.72  &   608.22  &   110.92  &  18.655 $\pm$ 0.106  &  17.725 $\pm$ 0.020  &  16.358 $\pm$ 0.012  &  15.357 $\pm$ 0.016  &  14.978 $\pm$ 0.016  &  ST \\
2013-11-12.67  &   609.17  &   111.87  &         ---          &  17.700 $\pm$ 0.021  &  16.379 $\pm$ 0.014  &  15.358 $\pm$ 0.017  &  15.004 $\pm$ 0.014  &  ST \\
2013-11-14.65  &   611.15  &   113.85  &         ---          &  17.764 $\pm$ 0.031  &  16.405 $\pm$ 0.011  &  15.402 $\pm$ 0.010  &  15.031 $\pm$ 0.013  &  ST \\
2013-11-19.69  &   616.19  &   118.89  &  18.515 $\pm$ 0.133  &  17.865 $\pm$ 0.023  &  16.493 $\pm$ 0.015  &  15.480 $\pm$ 0.016  &  15.133 $\pm$ 0.019  &  ST \\
2013-11-23.69  &   620.19  &   122.89  &  19.144 $\pm$ 0.408  &  17.945 $\pm$ 0.021  &  16.533 $\pm$ 0.009  &  15.529 $\pm$ 0.010  &  15.203 $\pm$ 0.011  &  ST \\
2013-11-24.62  &   621.12  &   123.82  &  18.973 $\pm$ 0.128  &  17.911 $\pm$ 0.019  &  16.552 $\pm$ 0.012  &  15.544 $\pm$ 0.015  &  15.205 $\pm$ 0.016  &  ST \\
2013-12-06.72  &   633.22  &   135.92  &  19.292 $\pm$ 0.171  &  18.113 $\pm$ 0.028  &  16.771 $\pm$ 0.014  &  15.719 $\pm$ 0.016  &  15.420 $\pm$ 0.017  &  ST \\
2013-12-08.73  &   635.23  &   137.93  &  19.286 $\pm$ 0.175  &  18.139 $\pm$ 0.018  &  16.815 $\pm$ 0.017  &  15.766 $\pm$ 0.022  &  15.486 $\pm$ 0.024  &  ST \\
2013-12-09.69  &   636.19  &   138.89  &         ---          &  18.167 $\pm$ 0.022  &  16.832 $\pm$ 0.011  &  15.779 $\pm$ 0.017  &  15.488 $\pm$ 0.017  &  ST \\
2013-12-10.61  &   637.11  &   139.81  &         ---          &  18.209 $\pm$ 0.034  &  16.863 $\pm$ 0.019  &  15.796 $\pm$ 0.020  &  15.490 $\pm$ 0.022  &  ST \\
2013-12-14.74  &   641.24  &   143.94  &         ---          &  18.015 $\pm$ 0.093  &  16.892 $\pm$ 0.034  &  15.856 $\pm$ 0.020  &  15.597 $\pm$ 0.023  &  ST \\
2013-12-15.63  &   642.13  &   144.83  &         ---          &  18.223 $\pm$ 0.041  &  16.974 $\pm$ 0.019  &  15.914 $\pm$ 0.025  &  15.603 $\pm$ 0.026  &  ST \\
2013-12-16.70  &   643.20  &   145.90  &         ---          &  18.109 $\pm$ 0.053  &  16.943 $\pm$ 0.025  &  15.903 $\pm$ 0.019  &  15.596 $\pm$ 0.126  &  ST \\
2013-12-19.61  &   646.11  &   148.81  &         ---          &  18.249 $\pm$ 0.043  &  17.009 $\pm$ 0.015  &  15.932 $\pm$ 0.019  &  15.661 $\pm$ 0.023  &  ST \\
2013-12-24.62  &   651.12  &   153.82  &  19.474 $\pm$ 0.061  &  18.265 $\pm$ 0.027  &  17.138 $\pm$ 0.014  &  16.003 $\pm$ 0.015  &  15.743 $\pm$ 0.016  &  ST \\
2013-12-25.66  &   652.16  &   154.86  &         ---          &  18.321 $\pm$ 0.016  &  17.101 $\pm$ 0.010  &  16.012 $\pm$ 0.009  &  15.722 $\pm$ 0.012  &  ST,DFOT \\
2013-12-28.62  &   655.12  &   157.82  &  19.368 $\pm$ 0.058  &  18.325 $\pm$ 0.019  &  17.161 $\pm$ 0.009  &  16.041 $\pm$ 0.015  &  15.760 $\pm$ 0.016  &  DFOT \\
2013-12-29.59  &   656.09  &   158.79  &  19.436 $\pm$ 0.060  &  18.315 $\pm$ 0.024  &  17.180 $\pm$ 0.011  &  16.061 $\pm$ 0.010  &  15.791 $\pm$ 0.011  &  DFOT \\
2014-01-19.62  &   677.12  &   179.82  &         ---          &  18.676 $\pm$ 0.025  &  17.458 $\pm$ 0.011  &  16.370 $\pm$ 0.014  &  16.128 $\pm$ 0.015  &  ST \\
2014-01-25.62  &   683.12  &   185.82  &  19.703 $\pm$ 0.071  &  18.638 $\pm$ 0.013  &  17.526 $\pm$ 0.009  &  16.424 $\pm$ 0.011  &  16.164 $\pm$ 0.012  &  DFOT \\
2014-01-30.62  &   688.12  &   190.82  &  19.797 $\pm$ 0.596  &  18.785 $\pm$ 0.027  &  17.602 $\pm$ 0.014  &  16.501 $\pm$ 0.013  &  16.282 $\pm$ 0.015  &  ST \\
2014-01-31.58  &   689.08  &   191.78  &         ---          &  18.787 $\pm$ 0.030  &  17.618 $\pm$ 0.019  &  16.522 $\pm$ 0.017  &  16.273 $\pm$ 0.025  &  ST \\
2014-02-02.62  &   691.12  &   193.82  &         ---          &  18.813 $\pm$ 0.035  &  17.623 $\pm$ 0.031  &  16.546 $\pm$ 0.020  &  16.323 $\pm$ 0.024  &  ST \\
2014-02-17.59  &   706.09  &   208.79  &         ---          &  19.218 $\pm$ 0.079  &  17.814 $\pm$ 0.022  &  16.682 $\pm$ 0.012  &  16.470 $\pm$ 0.017  &  ST \\
    \hline

  \end{tabular}
  \\
  \textit{Swift}~UVOT photometry\\
  \begin{tabular}
  {c c r c c c c c c c l}
  \hline
    UT Date&JD&Phase$^{a}$     & $uvw2$& $uvm2$& $uvw1$& $uvu$& $uvb$& $uvv$&Tel$^{b}$ & \\
    (yyyy/mm/dd)&2456000+&(day)&(mag)& (mag)& (mag)& (mag)& (mag)& (mag)& /Inst & \\ \hline
2013-07-30.98  &   504.48   &    7.18    &  12.369 $\pm$ 0.040 &  12.023 $\pm$ 0.040 &  11.711 $\pm$ 0.039 &          ---        &          ---        &  12.689 $\pm$ 0.042  &  Swift   \\
2013-07-31.50  &   505.00   &    7.70    &  12.455 $\pm$ 0.040 &  12.097 $\pm$ 0.040 &  11.755 $\pm$ 0.039 &          ---        &          ---        &  12.614 $\pm$ 0.040  &  Swift   \\
2013-07-31.83  &   505.33   &    8.03    &  12.577 $\pm$ 0.035 &  12.204 $\pm$ 0.033 &  11.814 $\pm$ 0.032 &          ---        &          ---        &          ---         &  Swift   \\
2013-08-03.06  &   507.56   &   10.26    &  13.044 $\pm$ 0.037 &  12.695 $\pm$ 0.041 &          ---        &  11.675 $\pm$ 0.029 &  12.619 $\pm$ 0.029 &          ---         &  Swift   \\
2013-08-03.18  &   507.68   &   10.38    &  13.056 $\pm$ 0.035 &          ---        &          ---        &          ---        &  12.622 $\pm$ 0.029 &          ---         &  Swift   \\
2013-08-04.85  &   509.35   &   12.05    &  13.374 $\pm$ 0.040 &  13.155 $\pm$ 0.053 &          ---        &  11.812 $\pm$ 0.029 &  12.608 $\pm$ 0.029 &          ---         &  Swift   \\
2013-08-04.98  &   509.48   &   12.18    &  13.385 $\pm$ 0.037 &          ---        &          ---        &          ---        &          ---        &          ---         &  Swift   \\
2013-08-07.24  &   511.74   &   14.44    &  13.907 $\pm$ 0.041 &          ---        &  12.948 $\pm$ 0.042 &          ---        &          ---        &          ---         &  Swift   \\
2013-08-07.55  &   512.05   &   14.75    &  13.968 $\pm$ 0.050 &          ---        &          ---        &          ---        &          ---        &          ---         &  Swift   \\
2013-08-08.02  &   512.52   &   15.22    &  14.039 $\pm$ 0.052 &  14.058 $\pm$ 0.070 &  13.131 $\pm$ 0.038 &  12.185 $\pm$ 0.031 &  12.749 $\pm$ 0.029 &  12.477 $\pm$ 0.030  &  Swift   \\
2013-08-08.22  &   512.72   &   15.42    &  14.126 $\pm$ 0.045 &          ---        &          ---        &  12.266 $\pm$ 0.029 &          ---        &          ---         &  Swift   \\
2013-08-09.25  &   513.75   &   16.45    &  14.387 $\pm$ 0.055 &  14.305 $\pm$ 0.112 &  13.379 $\pm$ 0.041 &  12.333 $\pm$ 0.029 &  12.906 $\pm$ 0.029 &  12.535 $\pm$ 0.031  &  Swift   \\
2013-08-09.31  &   513.81   &   16.51    &          ---        &  14.406 $\pm$ 0.065 &          ---        &          ---        &          ---        &          ---         &  Swift   \\
2013-08-11.78  &   516.28   &   18.98    &  15.210 $\pm$ 0.118 &  15.114 $\pm$ 0.109 &  13.907 $\pm$ 0.052 &  12.659 $\pm$ 0.029 &  12.983 $\pm$ 0.029 &  12.581 $\pm$ 0.031  &  Swift   \\
2013-08-13.85  &   518.35   &   21.05    &  15.652 $\pm$ 0.082 &  15.964 $\pm$ 0.068 &  14.446 $\pm$ 0.059 &  12.982 $\pm$ 0.031 &  13.109 $\pm$ 0.029 &  12.599 $\pm$ 0.032  &  Swift   \\
2013-08-15.00  &   520.50   &   23.20    &  16.209 $\pm$ 0.090 &          ---        &  14.905 $\pm$ 0.069 &  13.308 $\pm$ 0.033 &  13.221 $\pm$ 0.030 &  12.573 $\pm$ 0.032  &  Swift   \\
2013-08-17.65  &   522.15   &   24.85    &  16.588 $\pm$ 0.098 &  17.109 $\pm$ 0.195 &  15.201 $\pm$ 0.072 &  13.602 $\pm$ 0.035 &  13.293 $\pm$ 0.030 &  12.656 $\pm$ 0.032  &  Swift   \\
2013-08-19.73  &   524.23   &   26.93    &  16.824 $\pm$ 0.105 &  17.554 $\pm$ 0.221 &  15.493 $\pm$ 0.076 &  13.964 $\pm$ 0.039 &  13.476 $\pm$ 0.030 &  12.692 $\pm$ 0.032  &  Swift   \\
2013-08-22.54  &   527.04   &   29.74    &  17.245 $\pm$ 0.120 &  18.047 $\pm$ 0.250 &  15.890 $\pm$ 0.075 &  14.338 $\pm$ 0.045 &  13.663 $\pm$ 0.032 &  12.816 $\pm$ 0.033  &  Swift   \\
2013-08-23.14  &   527.64   &   30.34    &  17.170 $\pm$ 0.117 &          ---        &  15.866 $\pm$ 0.083 &  14.366 $\pm$ 0.045 &  13.627 $\pm$ 0.031 &  12.892 $\pm$ 0.034  &  Swift   \\
2013-08-27.74  &   532.24   &   34.94    &  17.746 $\pm$ 0.146 &  18.569 $\pm$ 0.214 &  16.356 $\pm$ 0.095 &  14.844 $\pm$ 0.058 &  13.915 $\pm$ 0.033 &  12.965 $\pm$ 0.034  &  Swift   \\
2013-09-06.16  &   541.66   &   44.36    &  18.133 $\pm$ 0.124 &  19.137 $\pm$ 0.190 &  16.793 $\pm$ 0.084 &  15.573 $\pm$ 0.067 &          ---        &          ---         &  Swift   \\
2013-09-06.41  &   541.91   &   44.61    &          ---        &          ---        &          ---        &  15.674 $\pm$ 0.087 &  14.367 $\pm$ 0.036 &  13.231 $\pm$ 0.035  &  Swift   \\
2013-09-16.71  &   552.21   &   54.91    &  18.687 $\pm$ 0.158 &  19.486 $\pm$ 0.236 &  17.292 $\pm$ 0.096 &  16.229 $\pm$ 0.090 &  14.750 $\pm$ 0.039 &  13.470 $\pm$ 0.038  &  Swift   \\
2013-09-26.45  &   561.95   &   64.65    &  18.793 $\pm$ 0.166 &          ---        &  17.562 $\pm$ 0.123 &  16.585 $\pm$ 0.128 &  14.922 $\pm$ 0.042 &  13.604 $\pm$ 0.039  &  Swift   \\
2013-10-06.88  &   572.38   &   75.08    &  19.241 $\pm$ 0.231 &  19.883 $\pm$ 0.333 &  17.919 $\pm$ 0.133 &  17.055 $\pm$ 0.094 &          ---        &          ---         &  Swift   \\
2013-10-16.77  &   582.27   &   84.97    &  19.294 $\pm$ 0.247 &          ---        &  18.127 $\pm$ 0.170 &  17.286 $\pm$ 0.164 &  15.464 $\pm$ 0.055 &  14.029 $\pm$ 0.045  &  Swift   \\
2013-10-26.95  &   592.45   &   95.15    &          ---        &          ---        &  18.248 $\pm$ 0.190 &  17.514 $\pm$ 0.171 &          ---        &          ---         &  Swift   \\
2013-11-06.16  &   602.66   &  105.36    &          ---        &          ---        &          ---        &  18.774 $\pm$ 0.304 &          ---        &          ---         &  Swift   \\
2013-11-13.21  &   609.71   &  112.41    &          ---        &          ---        &  19.523 $\pm$ 0.351 &  19.058 $\pm$ 0.306 &          ---        &          ---         &  Swift   \\
2013-11-13.68  &   610.18   &  112.88    &          ---        &          ---        &          ---        &  18.816 $\pm$ 0.263 &  17.974 $\pm$ 0.210 &  16.512 $\pm$ 0.148  &  Swift   \\
2013-11-20.43  &   616.93   &  119.63    &          ---        &          ---        &          ---        &  18.977 $\pm$ 0.214 &  17.889 $\pm$ 0.090 &  16.718 $\pm$ 0.080  &  Swift   \\
2013-11-25.40  &   621.90   &  124.60    &          ---        &          ---        &          ---        &  19.162 $\pm$ 0.323 &          ---        &          ---         &  Swift   \\
2013-11-30.43  &   626.93   &  129.63    &          ---        &          ---        &  19.726 $\pm$ 0.313 &  19.342 $\pm$ 0.280 &  18.155 $\pm$ 0.101 &  16.834 $\pm$ 0.082  &  Swift   \\
2013-12-09.75  &   636.25   &  138.95    &          ---        &          ---        &  19.807 $\pm$ 0.327 &  19.343 $\pm$ 0.274 &  18.196 $\pm$ 0.102 &  16.928 $\pm$ 0.085  &  Swift   \\

  \hline
  \end{tabular}
\begin{flushleft}
  $^{a}$ with reference to the explosion epoch \EpEpoch\\
  $^{b}$ ST : 104-cm Sampurnanand Telescope, ARIES, India; DFOT : 130-cm Devasthal fast optical telescope, ARIES, India; HCT: 2m Himalyan Chandra Telescope, Hanle, India; Swift: \textit{Swift}~UVOT\\
  Note: Data observed within 5 Hrs, are represented under single epoch observation.
\end{flushleft}
\end{table*}

 \begin{eqnarray*}
 u &=& U + (7.800\pm0.005)  - (0.067\pm0.009) \cdot (U-B)\\
 b &=& B + (5.269\pm0.007)  - (0.060\pm0.009) \cdot (B-V)\\
 v &=& V + (4.677\pm0.004)  - (0.056\pm0.005) \cdot (B-V)\\
 r &=& R + (4.405\pm0.005)  - (0.038\pm0.010) \cdot (V-R)\\
 i &=& I + (4.821\pm0.006)  - (0.048\pm0.006) \cdot (V-I)
 \end{eqnarray*}

 \noindent
 where $u$, $b$, $v$, $r$ and $i$ are instrumental magnitudes corrected for
 time, aperture and airmass; $U$, $B$, $V$, $R$ and $I$ are standard magnitude.
 The standard-deviation of the difference between the calibrated and the standard magnitudes of the observed Landolt stars are
 found to be $\sim$ 0.03 mag in $U$, $\sim$ 0.02 mag in $BR$ and $\sim$ 0.01 mag in $VI$. The transformation coefficients
 were then used to generate eight local standard stars in the field of \sn, which are verified to be non-variable and have brightness similar to SN. These stars are identified in Fig.\ref{fig:snid} and the calibrated \textit{UBVRI} magnitudes are listed in Table~\ref{tab:photstar}. These selected eight local standards were further used to standardize the instrumental light curve of the SN. One of these stars (star B) is common to that used in the study by \cite{2014JAVSO.tmp..275R}, and its \textit{BVRI} magnitudes are found to lie within 0.03 mag of our calibrated magnitudes. {Our calibrated magnitudes for \sn\ are also found to be consistent within errors to that presented in earlier studies of the event \citep{2014MNRAS.438L.101V,2014JAVSO.tmp..275R}.}
 The standard photometric magnitudes of \sn\ are listed in Table~\ref{tab:photsn}.

 This supernova was also observed with the Ultra-Violet/Optical Telescope \citep[UVOT;][]{2005SSRv..120...95R} in six bands (viz. uvw2, uvm2, uvw1, uvu, uvb, uvv)  on the Swift spacecraft \citep{2004ApJ...611.1005G}.   The UV photometry was obtained from the Swift Optical/Ultraviolet Supernova Archive\footnote{http://swift.gsfc.nasa.gov/docs/swift/sne/swift\_sn.html} (SOUSA; \citealp{2014Ap&SS.354...89B}).  The reduction is based on that of \citet{2009AJ....137.4517B}, including subtraction of the host galaxy count rates and uses the revised UV zeropoints and time-dependent sensitivity from \citet{2011AIPC.1358..373B}.  The UVOT photometry is listed in Table.~\ref{tab:photsn}.  The first month of UVOT photometry was previously presented by  \cite{2014MNRAS.438L.101V}.

\subsection{Spectroscopy} \label{sec:obs.spec}

Spectroscopic observations have been carried out at 10 phases during 12 to 125d. Out of these, nine epochs of low resolution spectra are obtained from Himalaya Faint Object Spectrograph and Camera (HFOSC) mounted on 2.0m HCT.
Spectroscopy on the HCT/HFOSC was done using a slit width of 1.92 arcsec, and grisms with resolution   $ \lambda/\Delta\lambda = 1330$ for Gr7 and $2190$ for Gr8, and bandwidth coverage of $0.38 - 0.64$  $\mu m$ and $0.58 - 0.84$ $\mu m$ respectively.
One high resolution spectrum is obtained from the ARC Echelle Spectrograph (ARCES) mounted on 3.5m ARC telescope located at Apache Point Observatory (APO).
ARCES is a high resolution cross-dispersion echelle spectrograph, the spectrum is recorded in 107 echelle orders covering a wavelength range of $\lambda\sim$  0.32-1.00\mum, at resolution of $R\sim31500$ \citep{2003SPIE.4841.1145W}. Summary of spectroscopic observations is given in Table.~\ref{tab:speclog}.

\begin{table*}
\centering
  \caption{Summary of spectroscopic observations of \sn. The spectral observations are made at 10
           phases during 12d to 125d.}
  \label{tab:speclog}
  \begin{tabular}{lc cc cc}
    \hline
    UT Date       &JD      &Phase$^{a}$&Telescope$^{c}$&    Range$^{b}$& Exposure \\
    (yy/mm/dd.dd) &2456000+&(days)     &               & \mum          & (s)      \\ \hline
2013-08-04.86  & 509.36  & ~12.1 &  HCT  &   0.38-0.68  &   900  \\
2013-08-27.76  & 532.26  & ~35.0 &  HCT  &   0.38-0.68  &  1200  \\
               &         &       &  HCT  &   0.58-0.84  &  1200  \\
2013-09-03.90  & 539.40  & ~42.1 &  HCT  &   0.38-0.68  &  1500  \\
               &         &       &  HCT  &   0.58-0.84  &  1500  \\
2013-09-29.78  & 565.28  & ~68.0 &  HCT  &   0.38-0.68  &  1800  \\
               &         &       &  HCT  &   0.58-0.84  &  2400  \\
2013-10-02.89  & 568.39  & ~71.1 &  HCT  &   0.38-0.68  &  1500  \\
               &         &       &  HCT  &   0.58-0.84  &  1500  \\
2013-10-11.28  & 576.78  & ~79.5 &  APO  &   0.32-1.00  &  1200  \\
2013-10-27.87  & 593.37  & ~96.1 &  HCT  &   0.38-0.68  &  2400  \\
2013-10-28.79  & 594.29  & ~97.0 &  HCT  &   0.58-0.84  &  2400  \\
2013-11-09.65  & 606.15  & 108.9 &  HCT  &   0.38-0.68  &  2100  \\
               &         &       &  HCT  &   0.58-0.84  &  3900  \\
2013-11-25.75  & 622.25  & 125.0 &  HCT  &   0.38-0.68  &  2400  \\
               &         &       &  HCT  &   0.58-0.84  &  2400  \\
    \hline
  \end{tabular}
\begin{flushleft}
  $^{a}$ With reference to the adopted explosion time \EpEpoch\\
  $^{b}$ For transmission $\ge$50\%\\
  $^{c}$ HCT : HFOSC on 2 m Himalyan Chandra Telescope, India; APO : Echelle spectrograph on 3.5 m ARC telescope at Apache Point Observatory, U.S. \\
  $^{d}$ At 0.6 \mum\\
\end{flushleft}
\end{table*}

Spectroscopic data reduction was done under the \iraf\ environment. Standard reduction procedures are followed for bias subtraction and flat fielding. Cosmic ray rejections are done using a Laplacian kernel detection algorithm for spectra, L.A.Cosmic \citep{2001PASP..113.1420V}. One dimensional low resolution spectra were extracted using the \textsc{apall} task. Wavelength calibration was done using the \textsc{identify} task applied on FeNe and FeAr (for HCT) arc spectra taken during observation.
Wavelength calibration was crosschecked against the [\ion{O}{I}] $ \lambda5577 $ sky line in the sky spectrum, and it was found to lie within 0.3 to 4.5 \AA\ of the actual value.
Spectra were flux calibrated using \textsc{standard, sensfunc} and \textsc{calibrate} tasks in \iraf. For flux calibration, spectrophotometric standards were used which were observed on the same nights as the SN spectra were recorded. All spectra were tied to absolute flux scale using the observed flux from \textit{UBVRI} photometry of SN. To perform the tying, individual spectrum is multiplied by a wavelength dependent polynomial, which is convolved with \textit{UBVRI} filters and then the polynomial is tuned to match the convolved flux with observations. The one dimensional calibrated spectra were corrected for heliocentric velocity of host galaxy (658 \kms; Table~\ref{tab:host}) using \textsc{dopcor} task.

\section{Distance and Extinction} \label{sec:ext}

We adopt a distance of $ 9.57\pm0.70$ Mpc which is a mean value of four different distance estimation techniques used for \host, viz., 9.91 Mpc applying Standard Candle Method (\textsc{scm}) to SN 2003gd by \cite{2010ApJ...715..833O}; 10.19 Mpc using the Tully-Fisher method (\texttt{HyperLeda}\footnote{http://leda.univ-lyon1.fr/}); 9.59 Mpc using brightest supergiant distance estimate by \cite{2005MNRAS.359..906H}; and planetary nebula luminosity function distance  8.59 Mpc \citep{2008ApJ...683..630H}. Although for each of these methods number of distance estimates exists in literature, we tried to select only most recent estimates. \cite{2014JAVSO.tmp..275R} estimated a distance of $ 9.1\pm0.4 $ Mpc by applying Expanding Photosphere Method (\epm) to \sn, which we find consistent to that we adopted.

One of the most reliable and well accepted method for SNe line-of-sight reddening estimation is using the \Nai~D absorption feature. The equivalent width (EW) of \Nai~D doublet (\ldld~5890, 5896) is found to be correlated with the reddening, estimated from the tail color curves of type Ia SNe \citep{1990A&A...237...79B,2003fthp.conf..200T}. However, \cite{2011MNRAS.415L..81P} suggested that although \Nai~D EW is weakly correlated with \ebv, the EWs estimated from low resolution spectra is a bad estimator of \ebv. \cite{2012MNRAS.426.1465P} used a larger sample of data and presented a more precise and rather different functional form of the correlation than that was derived earlier. Our high resolution echelle spectra at 79.5d provided an excellent opportunity to investigate the line-of-sight extinction.

\begin{figure}
 \centering
  \includegraphics[width=\linewidth]{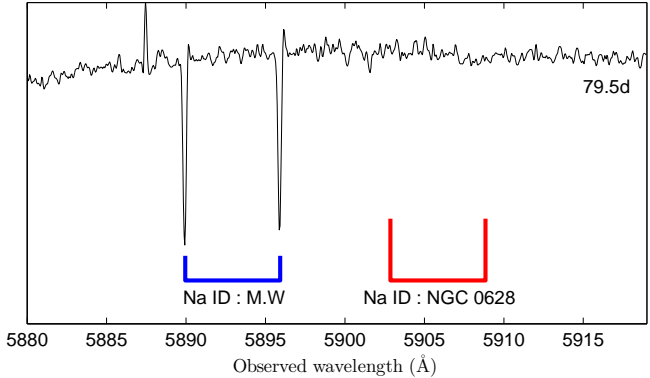}
 \caption{Echelle spectra at 79.5d showing the \Nai~D doublet for Milky-way while no impression for \host\ is detected.}
 \label{fig:naid.ech}
 \end{figure}

The resolved \Nai~D doublet for Milky-way is clearly visible in the high-resolution spectra (recorded on 79.5d) as shown in Fig.\ref{fig:naid.ech}. Whereas no impression of \Nai~D for \host\ is detected at the expected redshifted position relative to Milky-way. This indicates that the reddening due to host is negligible, only Galactic reddening will contribute to the total line of sight extinction. A similar conclusion has also been inferred by \cite{2014MNRAS.438L.101V} from their high resolution spectra obtained at 31d. Thus, we adopt a total $ \ebv=0.060\pm0.001 $ mag, which is entirely due to Galactic reddening \citep{2011ApJ...737..103S} and assuming total-to-selective extinction at V band as $ R_V=3.1 $, it translates into $ A_V=0.185\pm0.004 $ mag.

\section{Light curve} \label{sec:lc}

 \subsection{Light curve evolution and comparison} \label{sec:lc.app}

\begin{figure}
\centering
\includegraphics[width=\linewidth]{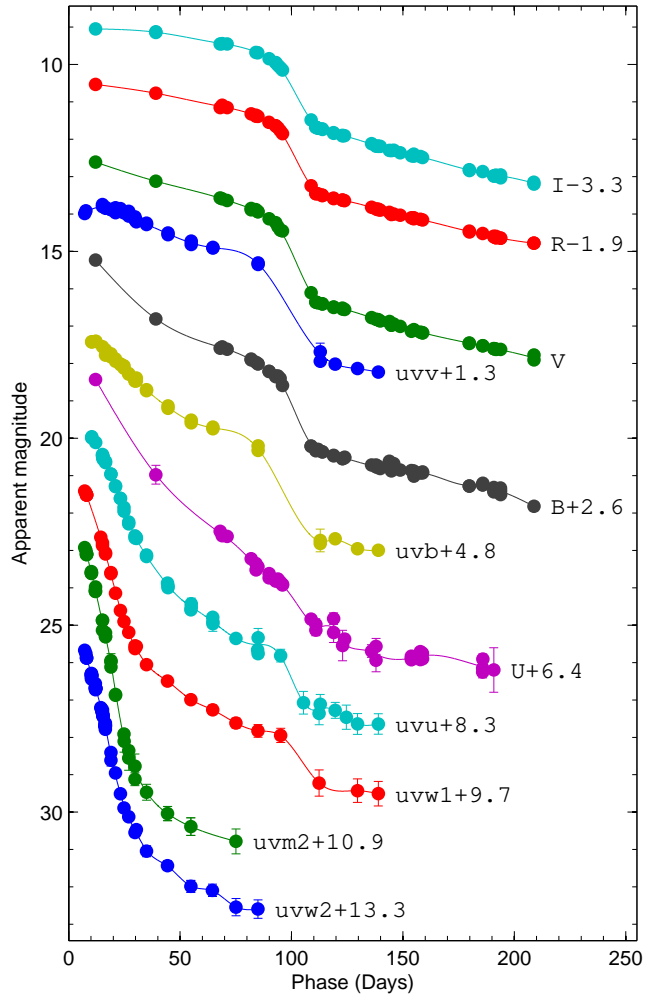}
\caption{The photometric light curves in Johnson-Cousins \textit{UBVRI} and \textit{Swift}~UVOT bands. The light curves are vertically shifted for clarity. The line joining the data points of light curves is for visualization purpose only.}
\label{fig:lc.app}
\end{figure}

The optical light curves of \sn\ in \textit{UBVRI} and six UVOT bands are shown in Fig.~\ref{fig:lc.app}. \textit{UBVRI} photometric observations were done at 38 phases during 12 to 209d (from plateau to nebular phase). The duration of plateau phase is sparsely covered, while denser follow-up initiated after 68d. The plateau phase lasted until $ \sim85 $d with an average decline rate of 6.60, 3.57, 1.74, 1.07 and 0.74 mag (100 d)$ ^{-1} $ in \textit{UBVRI} bands respectively. Since 95d, the light curve declines very fast until  115d, after which it settles to a relatively slow declining nebular phase. During this phase the decline rates for \textit{UBVRI} bands are 0.98, 1.22, 1.53, 1.42 and 1.55 mag (100 d)$ ^{-1} $ respectively.

\sn\ has been also observed by \textit{Swift}~UVOT at 35 phases during 7 to 139d. The UVOT \textit{UV} band light curves declines steeply during the first 30d at a rate of 0.182, 0.213, 0.262  mag d$ ^{-1} $ in \textit{uvw1, uvw2} and \textit{uvm2} bands respectively, thereafter settling into a slow declining  phase until it reaches the end of plateau.

\sn\ experience a steeper plateau decline than that observed for SN 1999em \citep{2002AAS...201.2303L}, SN 1999gi \citep{2002AJ....124.2490L}, SN 2012aw \citep{2013MNRAS.433.1871B} and SN 2013ab \citep{2015arXiv150400838B}. For example, SN 2012aw plateau declines at a rate of 5.60, 1.74, 0.55 mag (100 d)$ ^{-1} $ in $ UBV $-bands, similarly for SN 2013ab decline rates in \textit{UBVRI} are 7.60, 2.72, 0.92, 0.59 and 0.30 mag (100 d)$ ^{-1} $ and 0.169, 0.236, 0.257 mag d$ ^{-1} $ in UVOT \textit{uvw1, uvw2} and \textit{uvm2} bands (during first 30d).

\begin{figure*}
\centering
\includegraphics[width=16cm]{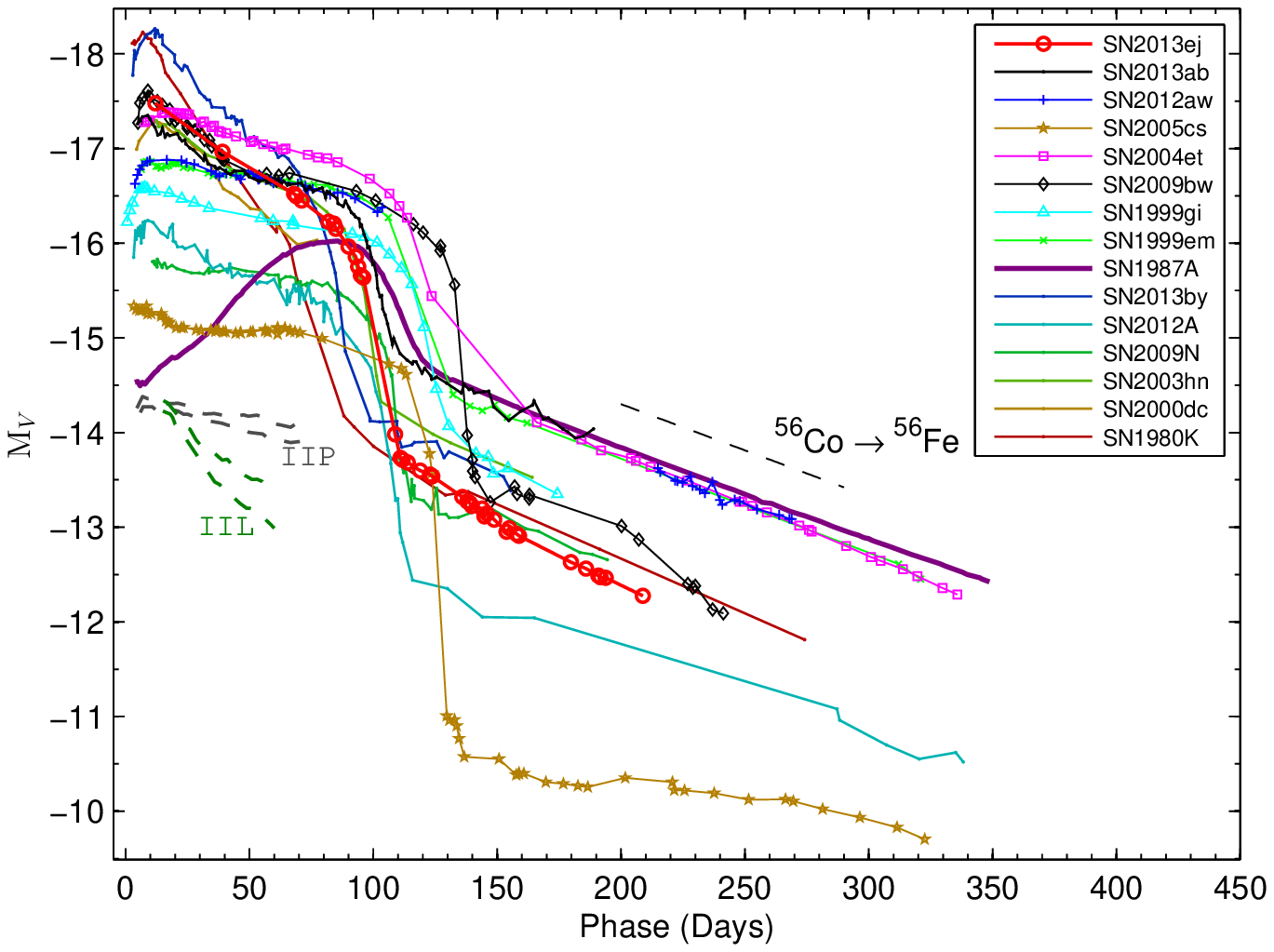}
\caption{
The M$_{V}$ light curve of \sn\ is compared with other type II SNe.
The exponential decline of the tail light curve following the radioactive decay law for \cobalt$ \rightarrow $\iron\ is shown with a dashed line. On the bottom left side, pair of dotted lines in each gray and green colors represent the slope range for type IIP and IIL SNe templates as given by \citet{2014MNRAS.445..554F}. The adopted explosion time in JD-2400000,
         distance in Mpc, \ebv\ in mag and
  the reference for observed V-band magnitude, respectively, are :
  SN 1980K  -- 44540.5, 5.5, 0.30;  \citet{1982A&A...116...35B}, NED database;
  SN 1987A  -- 46849.8, 0.05, 0.16;  \citet{1990AJ.....99.1146H};
  SN 1999em -- 51475.6, 11.7, 0.10; \citet{2002PASP..114...35L,2003MNRAS.338..939E};
  SN 1999gi -- 51522.3, 13.0, 0.21; \citet{2002AJ....124.2490L};
  SN 2000dc -- 51762.4, 49.0, 0.07; \citet{2014MNRAS.445..554F}, NED database;
  SN 2003hn -- 52866.5, 17.0, 0.19; \citet{2009AJ....137...34K,2014ApJ...786...67A};
  SN 2004et -- 53270.5, 5.4, 0.41; \citet{2006MNRAS.372.1315S};
  SN 2005cs -- 53549.0, 7.8, 0.11; \citet{2009MNRAS.394.2266P};
  SN 2009N  -- 54848.1, 21.6, 0.13; \citet{2014MNRAS.438..368T};
  SN 2009bw -- 54916.5, 20.2, 0.31; \citet{2012MNRAS.422.1122I};
  SN 2012A  -- 55933.5, 9.8, 0.04; \citet{2013MNRAS.434.1636T};
  SN 2012aw -- 56002.6, 9.9, 0.07; \citet{2013MNRAS.433.1871B};
  SN 2013ab -- 56340.0, 24.0, 0.04; \citet{2015arXiv150400838B};
  SN 2013by -- 56404.0, 14.8, 0.19; \citet{2015arXiv150106491V}.}
\label{fig:lc.abs}
\end{figure*}

{The absolute \textit{V}-band ($ M_V $) light curve of \sn\ is plotted in Fig.~\ref{fig:lc.abs} and is compared with other well studied type II SNe (after correcting for extinction and distance).
In Table~\ref{tab:slopendrop} we list the plateau slope of all compared type II events. The comparison shows that the decline rate of \sn\ during this phase is highest (1.74 mag (100 d)$ ^{-1} $) among most other SNe, except three type IIL SNe 1980K, 2000dc and 2013by, where SN 1980 is among the very first observed prototypical type IIL event. The early plateau ($ <40 $d) light curve of \sn\ is identical to SN 2009bw. However, unlike most other IIP SNe, e.g. 2009bw and 2013ab, which becomes flatter during late plateau, \sn\ continues to decline almost at a steady rate until the end of plateau ($ \sim $ 85d). The mid-plateau $ M_V=-14.7 $ mag for \sn, which places it in the class of normal luminous type II events.
\sn\ is comparable with fast declining and short plateau SNe in the sample of \cite{2014ApJ...786...67A}.
Following the plateau phase, $ V $-band light drops very fast to reach slow declining nebular phase (1.53 mag (100 d)$ ^{-1} $), which is powered by the radioactive decay of $ ^{56} $Co to $ ^{56} $Fe. The fall of  $ M_V $ during the plateau nebular transition is $ \sim 2.4$ mag, which is on the higher side of the compared events. The closest comparison is SNe 2009bw and 2012A which exhibits a drop of $ \sim $2.4 mag and $ \sim $2.5 mag respectively.} This also indicates low amount of \nickel\ mass synthesized during the explosion which we shall further discuss in the next section.

\begin{table}
  \centering
  \caption{Parameters estimated from {\it V} band light cruve}
  \label{tab:slopendrop}
  \begin{tabular}{lccc } \hline

       SN Name     &Plateau slope$^{a}$ &Transition drop$^{b}$&Transition time$^{c}$\\
                   &mag (100 d)$ ^{-1} $&mag                  &days           \\ \hline
		SN1980K    & 3.63 $\pm$ 0.04    &  2.0$\pm$0.04       & 37  $\pm$  5  \\
		SN2000dc   &~2.56 $\pm$ 0.06$^i$&  --                 & --            \\
		SN2013by   & 2.01 $\pm$ 0.02    &  2.2$\pm$0.03       & 19  $\pm$  5  \\
		SN 2013ej  & 1.74 $\pm$ 0.08    &  2.4$\pm$0.02       & 21  $\pm$  3  \\
		SN2003hn   & 1.41 $\pm$ 0.04    &  2.0$\pm$0.04       & 19  $\pm$  4  \\
		SN2012A    & 1.12 $\pm$ 0.03    &  2.5$\pm$0.02       & 23  $\pm$  4  \\
		SN2009bw   & 0.93 $\pm$ 0.04    &  2.4$\pm$0.03       & 14  $\pm$  3  \\
		SN2004et   & 0.73 $\pm$ 0.02    &  2.1$\pm$0.04       & 27  $\pm$  6  \\
		SN2013ab   & 0.54 $\pm$ 0.02    &  1.7$\pm$0.02       & 25  $\pm$  2  \\
		SN2012aw   & 0.51 $\pm$ 0.02    &   --                & --            \\
		SN1999gi   & 0.47 $\pm$ 0.02    &  2.0$\pm$0.02       & 29  $\pm$  3  \\
		SN2005cs   & 0.44 $\pm$ 0.03    &  4.0$\pm$0.03       & 24  $\pm$  3  \\
		SN2009N    & 0.36 $\pm$ 0.03    &  2.0$\pm$0.04       & 26  $\pm$  3  \\
		SN1999em   & 0.31 $\pm$ 0.02    &  1.9$\pm$0.02       & 28  $\pm$  4  \\
 \hline
  \end{tabular}
\begin{flushleft}
Note: Objects are sorted in order of plateau slope.\\
  $^{a}$ Plateau slope during the linear decline phase, starting after first minima until plateau end.\\
  $^{b}$ Drop in magnitude during the plateau to nebular transition.\\
  $^{c}$ Duration of plateau to nebular transition.\\
  $^{i}$ Slope is calculated up to the available range of data, as plateau end is not observed.\\
\end{flushleft}
\end{table}

\begin{figure}
\centering
\includegraphics[width=\linewidth]{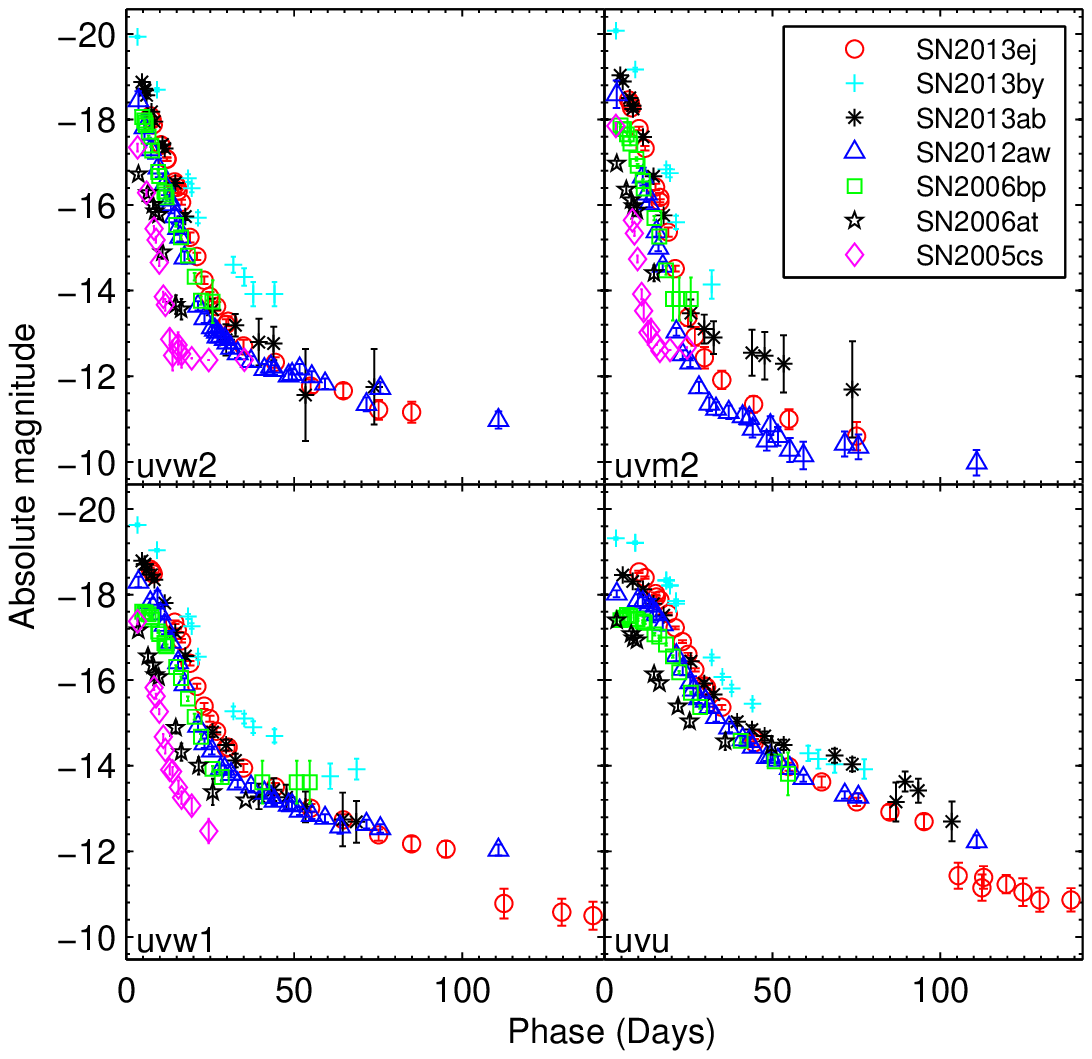}
\caption{Comparison of the \textit{Swift}~UVOT UV absolute light curves of \sn, with other well observed II SNe from UVOT. For the compared SNe, references for UVOT data, extinction and distance are: SN 2005cs -- \citet{2009AJ....137.4517B,2009MNRAS.394.2266P}, SN 2006at -- \citet{2009AJ....137.4517B}; Distance 65 Mpc; $ \ebv=0.031 $ mag \citep[only Galactic reddening][]{2011ApJ...737..103S}, SN 2006bp -- \citet{2008ApJ...675..644D}, SN 2012aw -- \citet{2013ApJ...764L..13B,2013MNRAS.433.1871B}, SN 2013ab -- \citet{2015arXiv150400838B}, SN 2013by -- \citet{2014Ap&SS.354...89B,2015arXiv150106491V}. Some late data points for SN 2013ab with large errors  has been omitted from the plot.}
\label{fig:uv.abs}
\end{figure}

\textit{Swift}~UVOT absolute magnitude light curves of \sn\ are shown in Fig.~\ref{fig:uv.abs} and compared with other well observed type II SNe. The sample is selected in such a way that SNe have at least a month of observations. Most SNe are not followed for more than a month by \textit{Swift}, mainly because of the large distances or high  extinction values. However, both these factors work in favor of \sn\, making it possible to have about four months of observations. Moreover, the location of the SN being in the outskirt of a spiral arm of \host, the background flux contamination is also negligible. The comparison shows that the \sn\ UV light curves are identical to SN 2012aw. \sn\ also shows a similar UV plateau trend as observed in SN 2012aw \citep{2013ApJ...764L..13B}, which is although expected but rarely detected for IIP/L SNe.

\begin{figure}
 \centering
 \includegraphics[width=8.75cm]{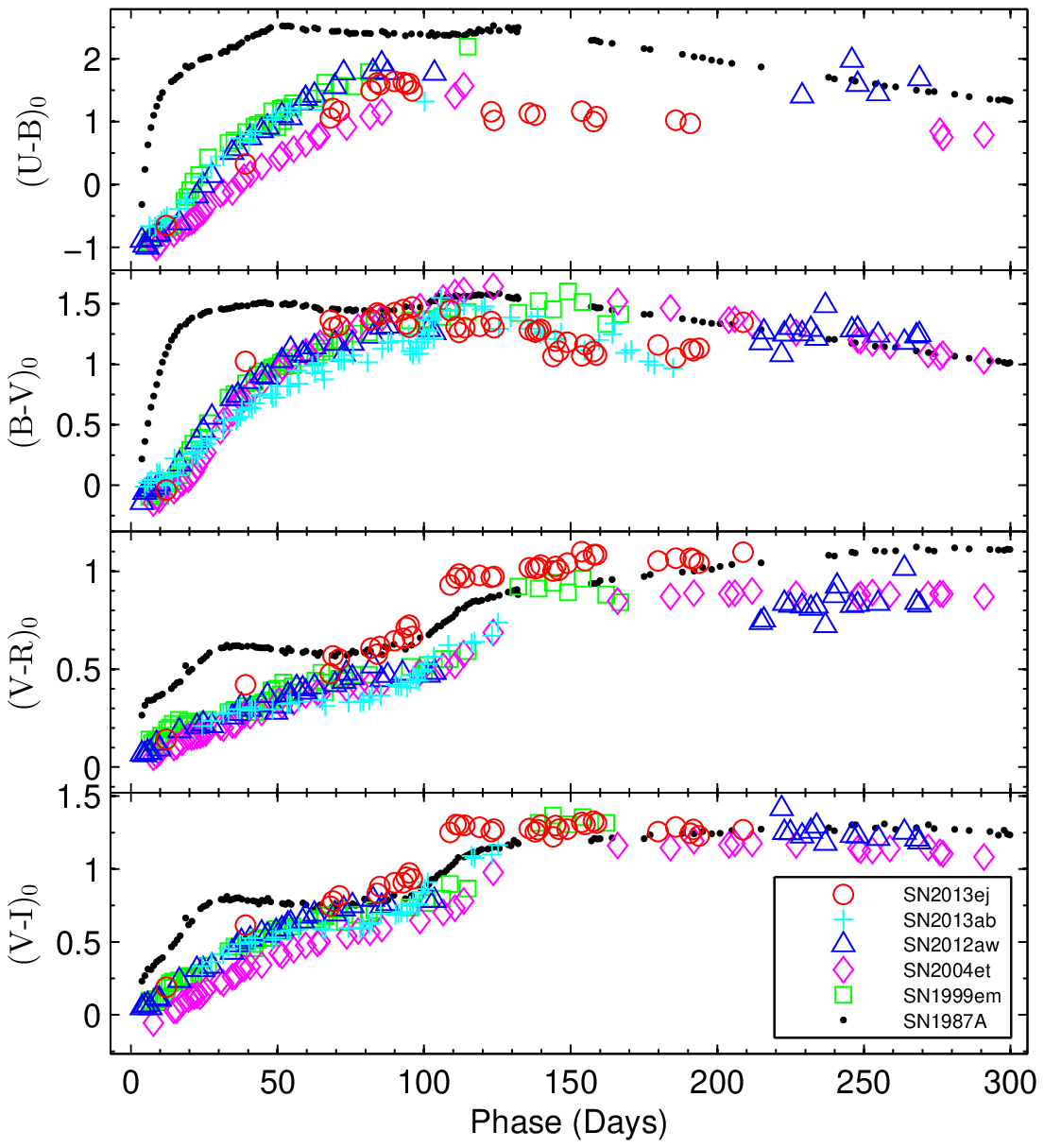}
 \caption{The intrinsic colors evolution of \sn\ is compared with other well-studied
          type IIP SNe 1987A, 1999em, 2004et, 2012aw and 2013ab. The reference for the data is same as in Fig.~\ref{fig:lc.abs}.}
 \label{fig:cc.abs}
 \end{figure}

Broadband color provides important information to study the temporal evolution of SN envelope. In Fig.~\ref{fig:cc.abs}, we plot the intrinsic colors \textit{U-B, B-V, V-R} and \textit{V-I} for \sn\ and compare its evolution with type II-pec SN 1987A, and type IIP SNe 1999em, 2004et, 2012aw and 2013ab. All the colors show generic signature of fast cooling ejecta until the end of plateau ($ \sim110 $d). With the start of the nebular phase it continues to cool at a much slower rate in \textit{V-I} and \textit{V-R} colors, whereas \textit{U-V} and \textit{B-V} shows a bluer trend. This is because, as the SN enters the nebular phase, the ejecta become depleted of free electrons, thereby making the envelope optically thin, and so unable to thermalize the photons from radioactive decay of $ ^{56} $Co to $ ^{56} $Fe.

 \subsection{Bolometric light-curve} \label{sec:lc.bol}

We compute the pseudo-bolometric luminosities following the method described in \cite{2013MNRAS.433.1871B}; which include SED integration over the semi-deconvolved photometric fluxes after correcting for extinction and distance. Supernova bolometric luminosities during early phases ($ \le30 $d) are dominated by ultraviolet fluxes, while after mid-plateau ($ \sim50 $d) UV contribution becomes insignificant as compared to optical counterpart  \citep[e.g., as seen in SNe 2012aw, 2013ab;][]{2013MNRAS.433.1871B,2015arXiv150400838B}. Similarly, during late phases $ >100 $d NIR becomes dominant over optical fluxes. However, during most of the light curve evolution, optical fluxes still provide significant contribution. We compute pseudo-bolometric luminosities in the wavelength range of \textit{U} to \textit{I} band (3335-8750\AA). We also computed UV-optical pseudo-bolometric light curve with wavelength starting from \textit{uvw2} band (wavelength range of 1606-8750\AA). The UV contribution enhances the luminosity significantly during early phases, whereas it is almost negligible after mid-plateau.

\begin{figure}
 \includegraphics[width=8.75cm]{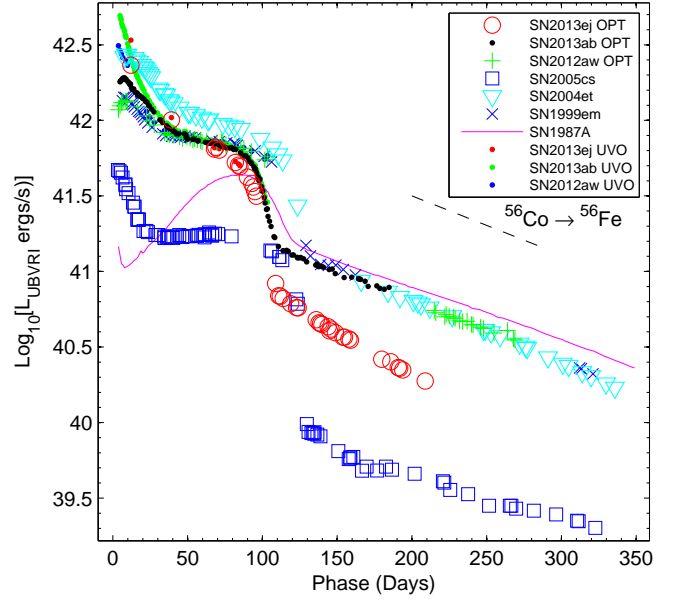}
 \caption{The \textit{UBVRI} bolometric light-curve of {\sn} is compared with other well studied supernovae. Light
 curves with added UVOT UV contributions are also shown for SNe 2013ej, 2013ab and 2012aw (labeled as UVO).
          The adopted values of distances, reddening and explosion time are same as in Fig.~\ref{fig:lc.abs}. The exponential decline of the tail light curve following the radioactive decay law is shown with a dashed line.}
 \label{fig:lc.bol}
 \end{figure}

In Fig.~\ref{fig:lc.bol}, we plot pseudo bolometric light curve for \sn\ and compare it with other SNe light curves computed using the same technique. We also include UV-optical bolometric light curve for SNe 2012aw and 2013ab along with SN 2013ej for comparison. Although the UV-optical light curve is initially brighter than the optical light curve, they completely coincide by the end of plateau phase (85d). It is evident from the comparison that  \sn\ experienced a steep decline during the plateau phase, but with a much shorter duration. This is consistent with the anti-correlation observed between plateau slope and duration for type II SNe \citep{1993A&A...273..106B,2014ApJ...786...67A}.
The UV-optical bolometric light decreases by 0.83 dex during plateau phase (from 12 to 85d), followed by an even faster drop by 0.76 dex in a short duration of 21 days (from 90 to 111d). Thereafter, the SN settles in a slow declining nebular phase. The tail luminosities are significantly lower than other normal luminosity IIP events, e.g., \sn\ luminosities are lower by $ \sim0.5 $ dex (at 200d) than that of type II SNe 1987A, 1999em, 2004et and 2012aw, but higher than subluminous events like SN 2005cs. Another noticeable dissimilarity of the tail light curve is its high decline rate. \sn\ tail luminosity declines at a rate of 0.55 dex 100 d$ ^{-1} $, which is much higher than that expected from radioactive decay of \cobalt\ to \iron. This is possibly because of inefficient gamma-ray trapping in the ejecta, and thus incomplete thermalization of the photons. We shall further explore this in \S\ref{modelling} in context of modeling the light curve.

 \subsection {Mass of nickel} \label{sec:lc.nick}

During the explosive nucleosynthesis of silicon and oxygen, at the time of shock-breakout in CCSNe, radioactive \nickel\ is produced. The nebular-phase light-curve is mainly powered by the radioactive decay of \nickel\ to \cobalt\ and \cobalt\ to \iron\ with half-life times
of 6.1d and 77.1d respectively
emitting $\gamma$-rays and
positrons. Thus the tail luminosity will be proportional to the amount of radioactive \nickel\ synthesized at the time of explosion.
We determine the mass of \nickel\ using following two methods.

For SN 1987A, one of the most well studied and well observed event, the mass of \nickel\ produced in the explosion has been estimated quite accurately, to be $ 0.075\pm0.005 $ \msun\ \citep{1996snih.book.....A}.
By comparing the tail luminosities of \sn\ and SN 1987A at similar phases, it is possible to estimate the \nickel\ mass for \sn.
In principle true bolometric luminosities (including UV, optical and IR) are to be used for this purpose, which are available for SN 1987A, whereas for \sn\ we have only UV and optical observations. Thus, in order to have uniformity in comparison, we used only the \textit{UBVRI} bolometric luminosities for both SNe and computed using the same method and wavelength range.
We estimate the tail \textit{UBVRI} luminosity at 175d, by making a linear fit over 155 to 195d, to be $ 2.90\pm0.43\times 10^{40} $ erg s$^{-1}$. Likewise, SN 1987A luminosity is estimated to be $ 9.60\pm0.06\times 10^{40} $ erg s$^{-1}$ at similar phase. Thus, the ratio of \sn\ to SN 1987A luminosity is $0.302\pm0.044$, which corresponds to a \nickel\ mass of $ 0.023\pm0.003 \msun$ for \sn.

Assuming the $\gamma$-photons emitted from radioactive decay of \cobalt\ thermalize the ejecta, \nickel\ mass can be independently estimated from the tail luminosity as described by \cite{2003ApJ...582..905H}.
 \begin{eqnarray*}
  M_{\rm Ni} = 7.866\times10^{-44} \times L_{t} \exp\left[ \frac{(t_{t}-t_{0})/(1+z)-6.1}{111.26}\right]\msun
 \end{eqnarray*}
where $t_{0}$ is the explosion time, 6.1d is the half-life time of \nickel\ and 111.26d is the e-folding time of the \cobalt\ decay. We compute tail luminosity $L_{t}$ at 6 epochs within 153 to 185d from the $ V $ band data corrected for distance, extinction and bolometric correction factor of $0.26 \pm 0.06$ mag during nebular phase \citep{2003ApJ...582..905H}. The weighted mean value of $L_{\rm t}$ is found to be $5.45\pm0.35\times10^{40}\,$\ergs corresponding to mean phase of 170d. This tail luminosity corresponds to a value of $M_{\rm Ni} =0.019\pm0.002$\msun.

We take the weighted mean of the estimated values from above two methods, and adopt a \nickel\ mass of $ 0.020\pm0.002 \msun$ for \sn.

\begin{figure}
\centering
\includegraphics[width=0.9\linewidth]{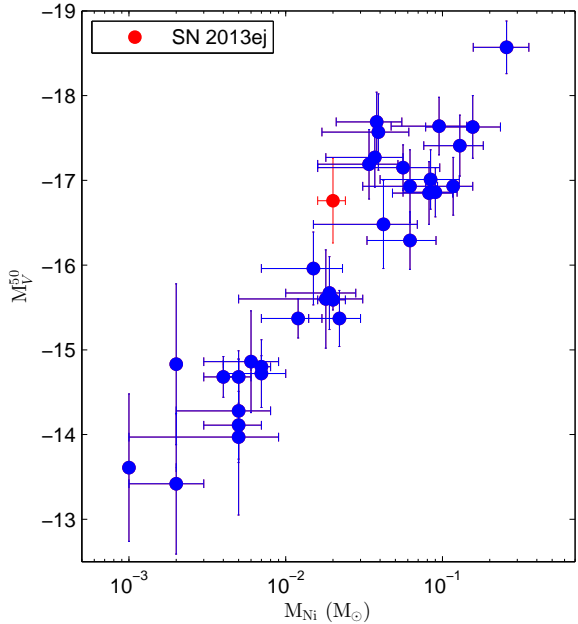}
\caption{The plot of absolute $ V $ band magnitude at 50 day versus \nickel\ mass for 34 type II SNe. Data taken from \citet{2003ApJ...582..905H} and \citet{2014MNRAS.439.2873S}. The position  of \sn\ on the correlation is shown with filled red circle.}
\label{fig:nicomp}
\end{figure}

{\cite{2003ApJ...582..905H} found a strong correlation between the \nickel\ mass and the mid plateau (at 50d) $ V $ band absolute magnitude for type II SNe and this correlation was further confirmed by \cite{2014MNRAS.439.2873S} specifically for low luminous events. Fig.~\ref{fig:nicomp} shows the correlation of mid plateau M$ _V $ versus \nickel\ mass for 34 events, including \sn. The SN lies within the scatter relation, but towards the lower mass range of \nickel\ than where most of the events cluster around (top right).}

\section{Optical spectra} \label{sec:sp}

\subsection{Key spectral features} \label{sec:sp.key}

\begin{figure*}
\centering
\includegraphics[width=14cm]{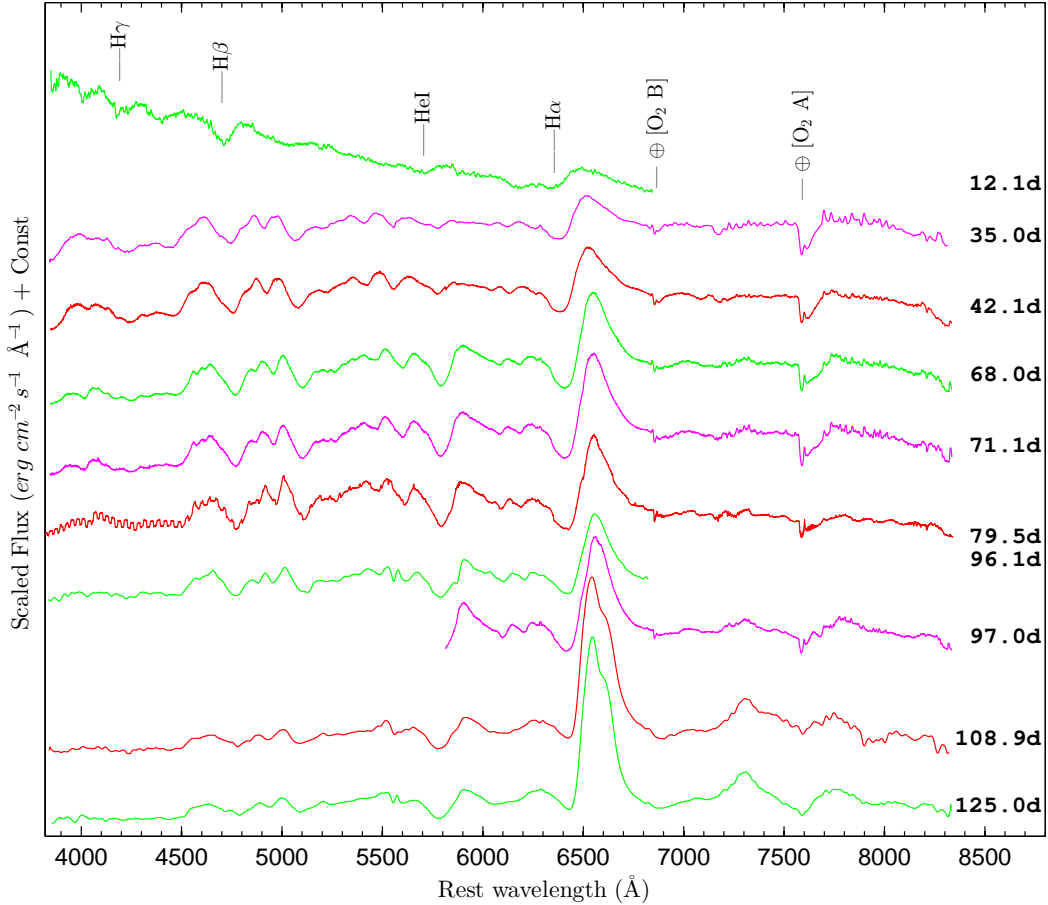}
\caption{The redshift corrected spectra of \sn\ are plotted for 10 phases during 12d to 125d. The prominent P-Cygni
         profiles of hydrogen (\ha, \hb, \hg) and helium (\Hei\ \ld5876) are marked. The telluric absorption features of O$ _2 $
         are marked with $ \oplus $, symbol. {Portion of spectra in extreme blue or red ends have low SNR. Individual spectra with with overall low SNR has been binned for better visualization.}}
\label{fig:sp.all}
\end{figure*}

\begin{figure}
 \centering
 \includegraphics[width=\linewidth]{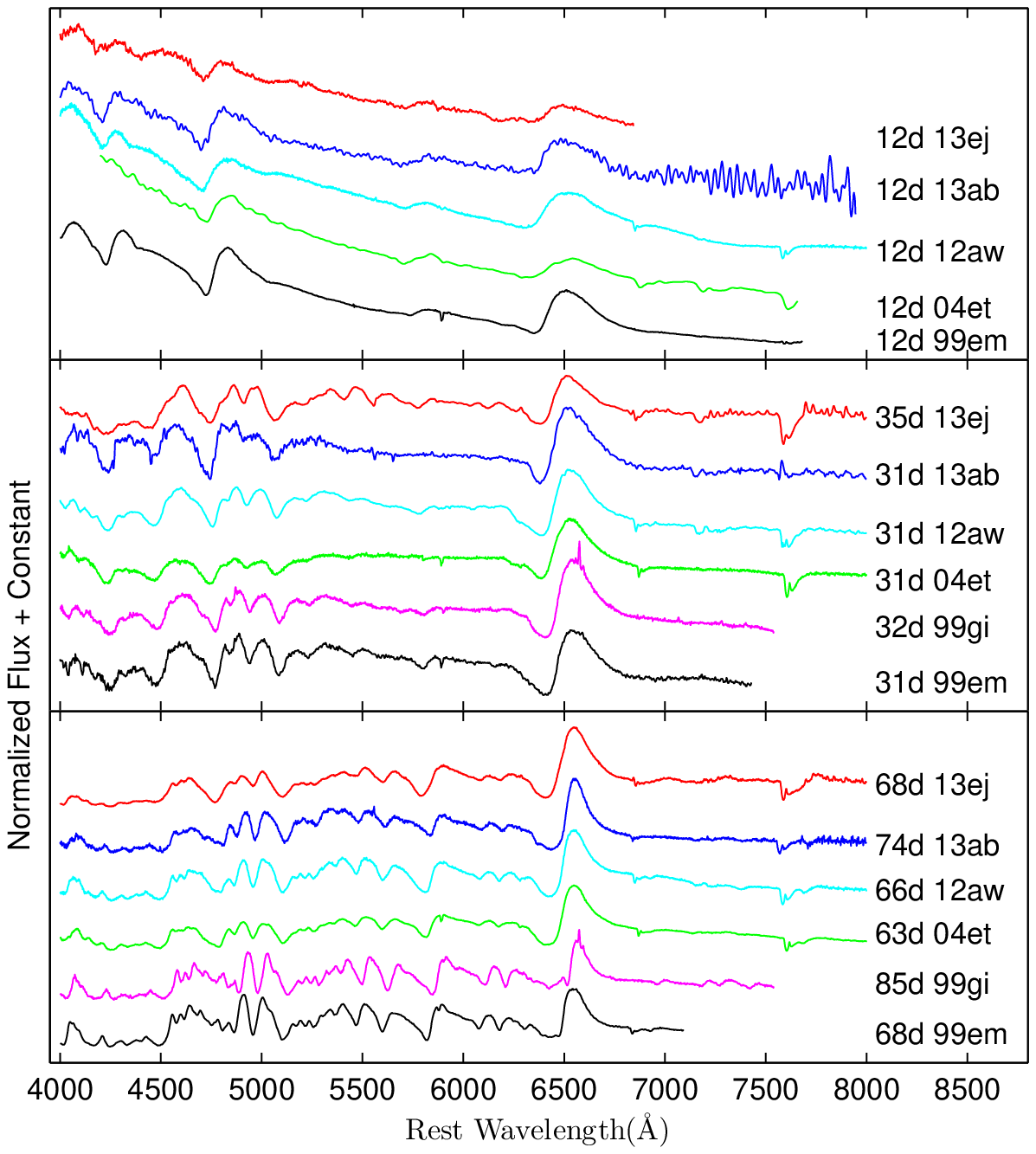}
 \caption{Comparison of early (12d) and  plateau (35d, 68d) phase spectra of {\sn} with
          other well-studied type IIP SNe 1999em \citep{2002PASP..114...35L}, 1999gi \citep{2002AJ....124.2490L},
          2004et \citep{2006MNRAS.372.1315S,2010MNRAS.404..981M}, 2012aw \citep{2013MNRAS.433.1871B} and 2013ab \citep{2015arXiv150400838B}. All
          comparison spectra are corrected for extinction and redshift (adopted values are same as in Fig.~\ref{fig:lc.abs}).}
 \label{fig:sp.lit}
 \end{figure}

The spectroscopic evolution of \sn\ is presented in Fig.~\ref{fig:sp.all}. Preliminary identifications of spectral features has been done as per previously studied type IIP SNe \citep[e.g.,][]{2002PASP..114...35L,2013MNRAS.433.1871B}. The spectrum at 12d shows broad \ha, \hb\ and \Hei\ features on top of a hot blue continuum. The 35d spectrum shows a relatively flat continuum with well developed features of \ha, \hb, \Feii\ along with blends of other heavier species \Tiii\ and \Baii. \Hei\ line is no longer detectable, instead \Nai~D features start to appear at similar location. The spectra from 35 to 80d represent the cooler photospheric phase, where the photosphere starts to penetrate deeper layers rich in heavier elements like \Feii\ and \Scii. During these phases we see the emergence and development of various other heavy atomic lines and their blends like \Tiii, \Baii, \Nai~D and \Caii.
Fig.~\ref{fig:sp.lit} shows the comparison of three plateau phase spectra, viz. 12, 35 and 68d with other well studied type IIP SNe at similar epochs. The comparison shows the spectra of \sn\ is broadly identical to others  in terms of observable line features and their evolution.
A notable feature during early spectrum (12d) is the dip on the bluer wing of \ha\ profiles near 6170 \AA\, which can be attributed as the \Siii\ feature.
{\cite{2013ATel.5275....1L} also identified this feature at $ \sim 9$d spectra of \sn\, however, due to unlikeliness of such a strong \Siii\ feature at such early epochs, a possiblity of non-standard red supergiant envelope or CSM interaction was suggested.}
However, such dips are detectable in 35 and 42d spectra, which we identify as \Siii\ feature in \synow\ modeling.

\begin{figure}
 \centering
 \includegraphics[width=8.5cm]{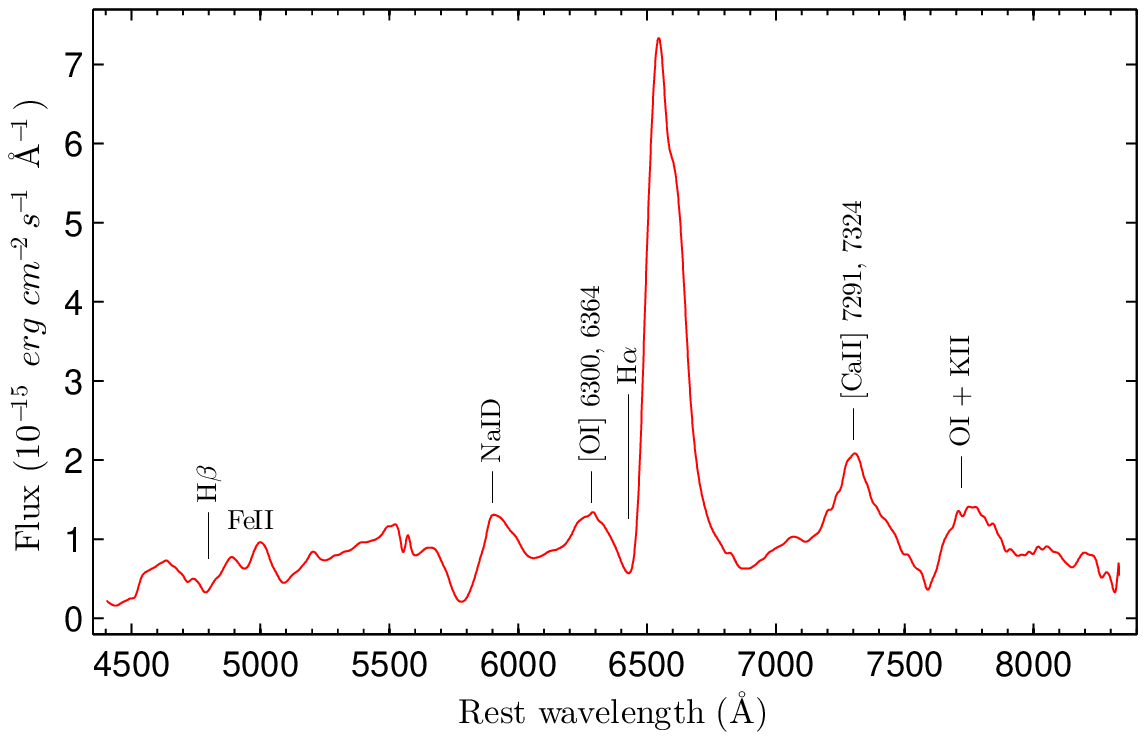}
 \caption{The nebular phase spectrum of \sn\ at 125d. Prominent emission and absorption features are marked and labeled.}
 \label{fig:sp.neb}
 \end{figure}

The spectra at 96 and 97d represents the plateau-nebular transition phase. Thereafter, spectra at 109 and 125d represents the nebular phase, where the ejecta has become optically thin. {These spectra shows the emergence of some emission features from forbidden lines of \Oia\ \ldld~6300, 6364 and \Caiia\ \ldld~7291, 7324\AA, as well as previously evolved permitted lines
 of \Hi, and the \Nai\ \ld5893 doublet (see Fig.~\ref{fig:sp.neb}).}

\cite{2014ApJ...786L..15G} found correlations between \ha\ absorption to emission strengths and light curve parameters, i.e. plateau slope and duration of optically thick phase. Following their selection criteria for choosing phase of SN spectra, i.e. ten days after start of recombination, we selected 42d spectrum as the closet available phase to the criteria. The \ha\ absorption to emission ratio of equivalent widths for \sn\ is found to be $ 0.23\pm0.02 $, the optically thick phase is $ \sim85 $d and \textit{B}-band late plateau (40 to 85d) slope is $ \sim0.27 $ mag (100 d)$ ^{-1} $. The correlation for optically thick phase duration is found to follow that presented by \cite{2014ApJ...786L..15G}. For the plateau slope, the correlation also hold true, but here \sn\ lies in the border line position of the scattered relation.
However, it may be noted that \ha\ profiles are possibly contaminated by high velocity features as we describe in next sections, which may result in deviation from correlation.

\subsection{\textsc{SYNOW} modelling of spectra} \label{sec:synow}
\sn\ spectra has been modeled with \synow\footnote{https://c3.lbl.gov/es/\#id22} \citep{1997ApJ...481L..89F,1999MNRAS.304...67F,2002ApJ...566.1005B} for line identification and its velocity estimation. \synow\ is a highly parametrized spectrum synthesis code which employs the Sobolev approximation to simplify radiation transfer equations assuming a spherically symmetric supernova expanding homologously. The strength of the \synow\ code is its capability to reproduce P-Cygni profiles simultaneously in synthetic spectra for a given set of atomic species and ionization states.
The applicability of \synow\ is well tested in various core-collapse SNe studies \citep[e.g.][]{2012MNRAS.422.1178I,2013MNRAS.433.1871B,2013ApJ...767...71M,2014ApJ...782...98B, 2014MNRAS.438..368T,2014ApJ...781...69M} for velocity estimation and analysis of spectral lines.

To model the spectra we tried various optical depth profiles (viz. gaussian, exponential and power law) with no significant difference among them, however we find exponential profile ($\tau\propto exp[-v/v_e]$) marginally better suited to match the absorption trough of observed spectra, where $v_{e}$ the e-folding velocity, is a fitted parameter. While modeling spectra, \Hi\ lines are always dealt as detached scenario. This implies the velocity of hydrogen layer is significantly higher and is thus detached from photospheric layer, close to which most heavier atomic lines form, as assumed in \synow\ code. As a consequence to this, the \ha\ lines in synthetic spectrum, which are highly detached, has flat topped emissions with blue shifted absorption counter parts.

\begin{figure*}
\centering
\includegraphics[width=0.8\linewidth]{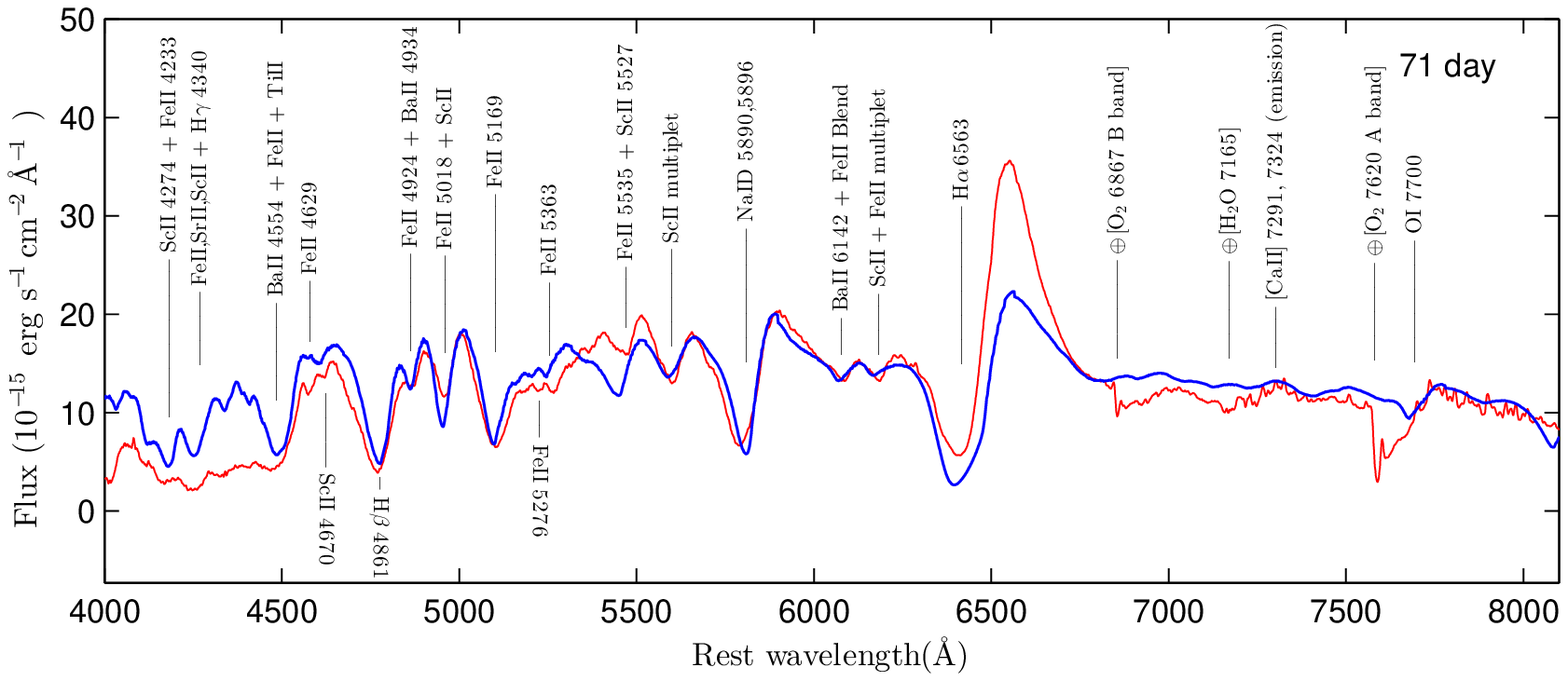}
\caption{\synow\ modelling of \sn\ spectrum at 71d. Model spectrum is shown with thick solid line (blue),
         while the observed one is shown with thin solid line (red). Observed fluxes are corrected for extinction.}
\label{fig:sp.synph}
\end{figure*}

\sn\ spectra are dereddened and approximate blackbody temperature is supplied in the model to match the spectral continuum. For early spectrum (12d), local thermodynamic equilibrium (LTE) assumption holds good and thus \synow\ could fit the continuum well, whereas at later epochs it fails to fit properly. The set of atomic species incorporated to generate the synthetic model spectrum are \Hi, \Hei; \Feii; \Tiii; \Scii; \Caii; \Baii; \Nai\ and \Siii. The photospheric velocity $ v_{\rm ph} $ is optimized to simultaneously fit the \Feii\ (\ldld~4924, 5018, 5169) P-Cygni profiles and \Hi\ lines are treated as detached. The optical depths and optical depth profile parameters, e-folding velocity are varied for individual species to fit respective line profiles. In Fig.~\ref{fig:sp.synph} we show the model fit of 71d spectrum. Most of the observable spectral features are reproduced well and are identified in the figure.

\begin{figure}
 \centering
 \includegraphics[height=7.75cm]{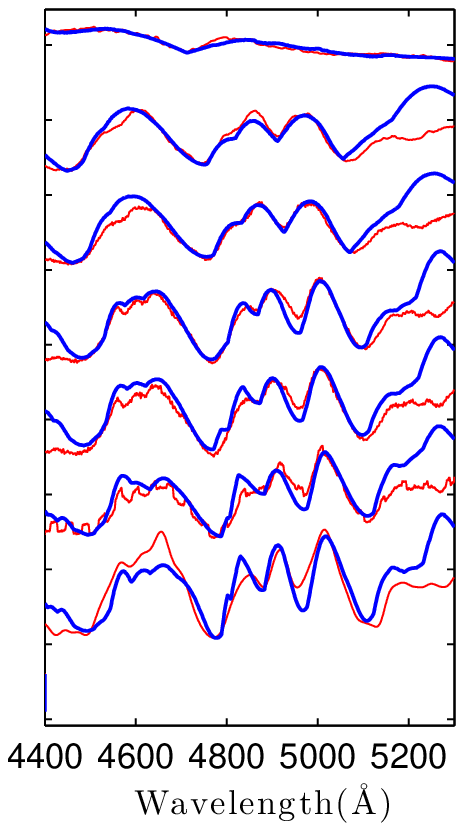}
 \includegraphics[height=7.75cm]{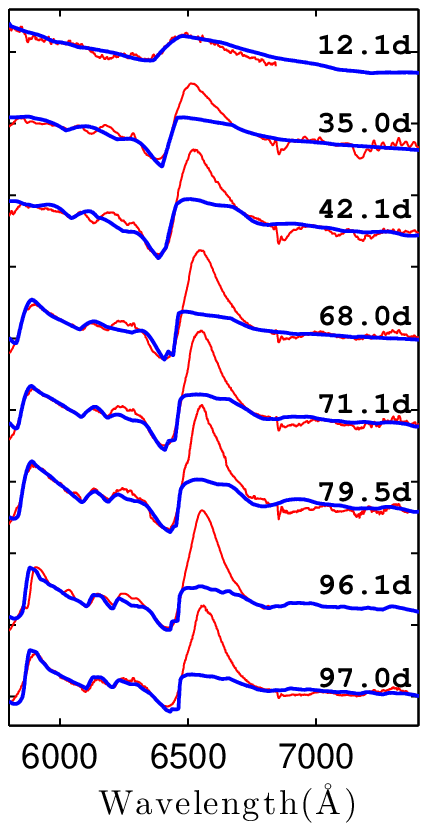}
 \caption{\synow\ modelling of \sn\ spectra at 8 phases during 12d to 97d for \hb, \Feii\ multiplet (left) and \ha\ (right) profiles. Model spectra are shown with thick solid line (blue), while the observed ones are shown with thin solid line (red).
 In the model, \Hi\ lines are treated as detached to fit the absorption troughs.
 Along with \Feii\ and \Hi; other ions (\Scii, \Baii, \ion{Si}{ii} and \Nai,
 \Tiii) are also incorporated in model to fit some weaker features, specially at later phases. In addition to this, high-velocity \Hi\ lines are also incorporated (42d onwards) to fit the extended \ha\ and \hb\ absorption troughs.
 The 97d spectrum do not have \hb\ and \Feii\ wavelength region, hence it is not shown here.
          }
 \label{fig:sp.synall}
 \end{figure}

Similarly all spectra during 12 to 97d are modeled with \synow. The model fits for \Feii\ (\ldld~4924, 5018, 5169), \hb\ and \ha\ spectral sections are shown in Fig.~\ref{fig:sp.synall}. The atomic species which are important to model these features are \Hi, \Feii, \Baii, \Tiii, \Scii\ and \Nai. In addition to these \Siii\ is also used to model the dips in the blue wing of \ha\ P-Cygni during 12 to 42d. While modeling the \ha\ and \hb\ profiles, \synow\ was unable to properly fit the broad and extended P-Cygni absorption troughs with single regular component. In order to fit these extended troughs, we invoke high-velocity (HV) component of \Hi. Although no separate dip is seen, possibly due to low spectral resolution and overlapping of broad P-Cygni profiles, the HV component can well reproduce the observed features in synthetic model spectrum. The implication and interpretation of these HV components are further discussed in \S\ref{sec:sp.vel}. The \synow-derived velocities for \Feii, \ha, \hb\ lines and corresponding HV components are listed in Table~\ref{tab:synow}. The nebular spectra during 109 to 125d have not been modeled primarily due to limitations of the LTE assumption of \synow, and also because nebular phase spectra are dominated by emission lines rather than P-Cygni profiles.

\begin{table*}
  \centering
  \caption{The line velocities
           of \ha, \hb, \Feii\ (\ldld~4924, 5018, 5169) and \Hei\ \ld5876 as estimated
           by modelling the observed spectra of \sn\ with \synow. \Feii\ or \Hei\ lines velocities are taken to represent photospheric velocity ($v_{\rm phm}$).}
  \label{tab:synow}

  \begin{tabular}{cccccccc } \hline
       UT Date     &Phase$^{a}$ & $v$(\Hei)   & $v$(\Feii)  & $v$(H$_\alpha$) & $v$(H$_\alpha$) HV$^b$& $v$(H$_\beta$) & $v$(H$_\beta$) HV$^b$\\
     (yyyy-mm-dd)  &   (day)    &$10^{3}$ \kms& $10^{3}$\kms& $10^{3}$\kms    & $10^{3}$\kms      & $10^{3}$\kms   & $10^{3}$\kms     \\ \hline
	 2013-08-04.86 &  12.1      &  8.8        &  --         &  9.6            &  --               &  9.7           &  --              \\
	 2013-08-27.76 &  35.0      &  --         &  6.7        &  7.9            &  --               &  6.6           &  --              \\
	 2013-09-03.90 &  42.1      &  --         &  5.8        &  7.2            &  8.5              &  5.4           &  6.4             \\
	 2013-09-29.78 &  68.0      &  --         &  3.6        &  5.8            &  7.4              &  4.0           &  5.8             \\
	 2013-10-02.89 &  71.1      &  --         &  3.3        &  5.4            &  7.3              &  3.8           &  5.8             \\
	 2013-10-11.28 &  79.5      &  --         &  3.3        &  5.2            &  6.3              &  3.6           &  4.8             \\
	 2013-10-27.87 &  96.1      &  --         &  2.7        &  4.9            &  6.3              &  3.5           &  4.8             \\
	 2013-10-28.79 &  97.0      &  --         &  2.7        &  4.8            &  6.3              &  --            &  --              \\
 \hline

  \end{tabular}
\begin{flushleft}
  $^{a}$ With reference to the time of explosion \EpEpoch\\
  $^{b}$ High velocity component used to fit the broad \ha\ and \hb\ profile.\\
\end{flushleft}
\end{table*}

\subsection{Evolution of spectral lines} \label{sec:sp.line}

Investigation of the spectral evolution sheds light on various important aspects of the SN, like interaction of ejecta
with the circumstellar material, geometrical distribution of expanding shell of ejecta and formation of dust during late time. SN spectra are dominated by P-Cygni profiles which are direct indicators of expansion velocities and they evolve with the velocity of photosphere. As ejecta expands and opacity decreases allowing photons to escape from deeper layers rich in heavier elements, we are able to see emergence and growth of various spectral lines.

\begin{figure}
 \centering
 \includegraphics[width=0.7\linewidth]{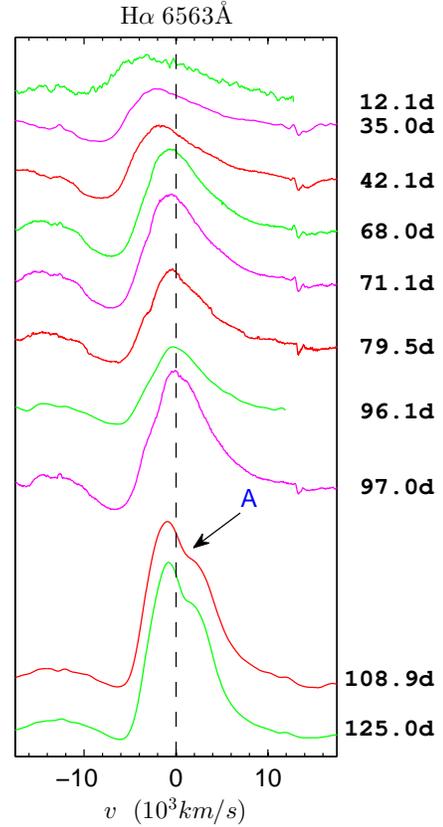}
 \caption{Evolution of \ha\ line profile at 10 phases during 12d to 125d.
 A zero-velocity line is plotted with a dashed line corresponding to the rest
          wavelength of \ha\ \ld6563.}
 \label{fig:sp.line}
 \end{figure}

To illustrate the evolution of \ha\ line,
in Fig.~\ref{fig:sp.line} partial region of spectra is plotted in velocity domain corresponding to rest wavelengths of \ha. At 12d broad P-Cygni profile (FWHM $ \sim9500 $ \kms) is visible which becomes narrower with time as the expansion slows down. The blue-shifted absorption troughs are direct estimator of expansion velocity of the associated line forming layer. The emission peaks are found to be blue-shifted (by $ \sim3200 $ \kms at 12d), which progressively decreases with decrease in expansion velocity and almost settling to zero velocity when the SN starts to enter nebular phase (97d). Such blue-shifted emission peaks, especially during early phases are generic features observable in SN spectra, e.g., SNe 1987A \citep{1987A&A...182L..29H}, 1998A \citep{2005MNRAS.360..950P}, 1999em \citep{2003MNRAS.338..939E}, 2004et \citep{2006MNRAS.372.1315S}, 2012aw \citep{2013MNRAS.433.1871B}, 2013ab \citep{2015arXiv150400838B}. These features are tied with the density structure of the ejecta, which in turn controls the amount of occultation of the receding part of ejecta, resulting in biasing of the emission peak \citep{2014MNRAS.441..671A}, which are not limited to \ha\ but applicable to all spectral lines. However, such a blue-shift is clearly detected for \ha\, whereas for most other lines, emission profiles are weak and peaks are contaminated by adjacent P-Cygni profiles. Detailed SN spectral synthesis code like \cmfgen\ \citep{2005ASPC..332..415D} is capable of reproducing such blue-shifted emission peaks.

As inferred from Fig.~\ref{fig:sp.lit}, the spectral evolution of \sn\ is almost identical to other typical IIP SNe. However, the comparison of 35 and 68d spectra indicates \Feii\ lines are somewhat under developed as compared to other SNe at similar phase. As seen in the 68d comparison, the \Feii\ (\ldld~4924, 5018, 5169) absorption dips are significantly weaker in comparison to that seen in other SNe.

Another prominent and unusual feature is seen in nebular spectra at 109d and 125d, on top of \ha\ emission, and the same is marked as feature A in Fig.~\ref{fig:sp.line}. This unusual dip is resulting into an apparent blue-shift of the emission peak, which is in fact larger than that seen in the last plateau spectra at 97d. Such evolution is unexpected and against the general trend of emission peak evolution in SNe. The low resolution of these spectra prohibits us from investigating this feature in detail.
{This feature can be split into two emission components, one redshifted at  1200 \kms\ and another blueshifted by 1300 \kms\ (see \S\ref{app:nebular_ha} for further explanation) with respect to \ha\ rest position. Such an asymmetric or double peaked \ha\ nebular emission has been observed in a number of SNe, e.g. SN 1999em \citep{2002PASP..114...35L} and SN 2004dj \citep{2005AstL...31..792C}. \cite{2002PASP..114...35L} identified such a dip or notch in \ha\ emission profile only during nebular phase of SN 1999em, which they suggested as possible ejecta-CSM interaction or asymmetry in line emitting region. In SN 2004dj, the asymmetry in nebular \ha\ spectra identified by \cite{2005AstL...31..792C} has been explained by bipolar distribution of \nickel\ with a spherical hydrogen envelope \citep{2006AstL...32..739C}.}

\subsection{Ejecta velocity} \label{sec:sp.vel}

Progenitor stars prior to explosion develop stratified layers of different elements, which are generally arranged in an elemental sequence, hydrogen being abundant in the outermost shell, whereas heavier metals like iron predominate at deeper layers. However at the time of shock breakout significant mixing of layers may occur. Spectral lines originating from different layers of the ejecta attains different characteristic velocities. Thus study of velocity evolution provides important clues to the explosion geometry and the characteristics of various layers. Evolution of photospheric layer is of special interest as it is directly connected to the kinematics and other related properties. Photosphere represents the layer of SN atmosphere where optical depth attains a value of $\sim~^2/_3 $ \citep{2005A&A...437..667D}. Due to complex mixing of layers and continuous recession of the recombination front, no single spectral line can  represent the true photospheric layer. During the plateau phase, \Feii\ or \Scii\ lines are the best estimator of photospheric velocity ($v_{\rm ph}$). In early phases when \Feii\ lines are not strongly detectable, the best proxy for $v_{\rm ph}$ is \Hei, or \hb\ \citep{2012MNRAS.419.2783T} in even earlier phases.

\begin{figure}
 \includegraphics[width=8.5cm]{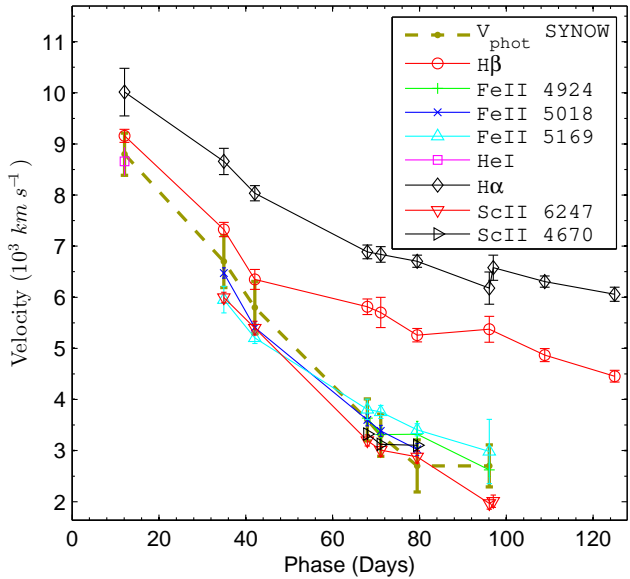}
 \caption{Velocity evolution of \ha, \hb, \Hei, \Scii\ and \Feii\ lines. The velocities are estimated using
          blueshift of the absorption minima. The expansion velocity of photosphere ($v_{\rm phm}$) estimated from
          \synow\ modeling of \Hei\ line at 12d and \Feii\ lines at later phases
          (see Table~\ref{tab:synow}) are also overplotted for comparison.}
 \label{fig:sp.velall}
 \end{figure}

Line velocities can either be estimated by directly locating the P-Cygni absorption troughs, as done using \textsc{splot} task of \iraf, or by by modeling the line profiles with velocity as one of the input, as we do in \synow. In Fig.~\ref{fig:sp.velall}, we plot the line velocities of \ha, \hb, \Feii\ (\ldld~4924, 5018, 5169) and \Scii\ (\ldld~4670, 6247), using the absorption minima method. It is evident that \Feii\ and \Scii\ line velocities are very close to each other and they are formed at deeper layers, whereas \ha\ and \hb\ line velocities are consistently higher at all phases as they form at larger radii. The \synow\ estimated photospheric velocities are also plotted for comparison, which is very close to the \Feii\ and \Scii\ velocities estimated from absorption minima method. Here the \synow-derived photospheric velocities are estimated by modelling \Hei\ line for 12d spectrum and \Feii\ lines for rest of the spectra. Velocities for various lines estimated using \synow\ are tabulated in Table~\ref{tab:synow}.

 Fig.~\ref{fig:sp.velph} shows the comparison of photospheric velocity of \sn\ with other well-studied type II
 SNe 1987A, 1999em, 1999gi, 2004et, 2005cs, 2012aw and 2013ab.
 For the purpose of comparison the absorption trough velocities have been used, taking the mean of \Feii\ line triplet, or \Hei\ lines at early phases where \Feii\ lines are not detectable. The velocity profile of \sn\
 is very similar to other normal IIP SNe 1999em, 1999gi, 2004et, 2012aw and 2013ab, on the other hand velocities of SN 2005cs and 1987A are significantly lower. The velocity profile of \sn\  is almost identical with SNe 2004et, 2012aw and 2013ab, whereas it is consistently higher than SNe 1999gi and 1999em by $ \sim800-900 $\kms.
 For comparison of \Hi\ (\ha\ and \hb) velocities, we have chosen all those events which are at least photometrically and spectroscopically similar to \sn. Comparison reveals that, H velocities during later phases (60-100 d) are consistently higher than all comparable events. SNe 2012aw and 2013ab, have photospheric velocities identical to \sn, but their H velocities are significantly lower by large values, e.g., for \sn\ the \ha\ velocity at 80d is higher by 1500 \kms\ and \hb\ is higher by 2400 \kms. Likewise, H velocities for SNe 1999em and 1999gi are even lower at similar phases. Although SN 2004et \Hi\ velocities are somewhat on higher end, they are still significantly less than those of \sn. It is also to be noted that, at 12d \sn\ \Hi\ velocities are consistent and similar to those of other normal SNe, but as it evolves these velocities decline relatively slowly, ultimately turning out into a higher velocity profile after $ \sim40 $d.

 \begin{figure}
 \centering
 \includegraphics[width=0.7\linewidth]{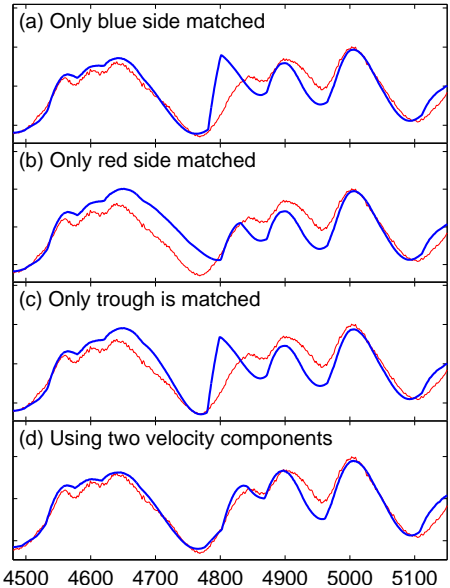}
 \caption{For 68d spectrum, the \hb\ profile is fitted using \synow\ with various velocity components. (a) The fit only with a single high velocity component to match the blue wing of the absorption dip, (b) with a single low velocity component to match the red wing, (c) with single velocity to only fit the trough, (d) with two velocity components to fit entire absorption profile.}
 \label{fig:synow.hvlv}
 \end{figure}

\subsection{High velocity components of \Hi\ and CSM interaction}
\label{sec:hvcsm}
As discussed in \S\ref{sec:synow}, the broad and extended \ha\ or \hb\ absorption profiles are not properly reproduced using single \Hi\ velocity component in \synow, and those profiles can only be fitted by incorporating a high-velocity (HV) components along with the regular one.
{Fig.~\ref{fig:synow.hvlv} shows the comparison of \synow\ fits for 68d \hb\ profile with various single velocity as well as for combined two velocity components. A single velocity component at 5600 \kms\ can match the blue wing well and partially the trough, whereas, it does not match the red side at all. Similarly, with a single velocity component at 4000 \kms\  can partially match the red slope of the trough, but does not include the trough as well as the extended blue wing. By only matching the trough position, the model fits for a single velocity of 5300 \kms, which does not fit either of the blue or red wing. Even-though the `detachment' of \Hi\ from photosphere in \synow\ model makes the fit of red wing worse by steepening it further, but it is still conclusive that none of these single velocity component can properly reproduce the absorption profile. It is only by including two velocity components together in the model could reproduce the entire \hb\ profile. Such a scenario start to appear from 42d spectrum which only  becomes stronger as the line evolves until 97d. The \ha\ troughs are also reproduced in a similar fashion.} {However, it may be noted, that such an extended \Hi\ feature may also be explained as a possible outcome of a different (complex and extended) density profile which \synow\ can not reproduce.}

\begin{figure}
\centering
\includegraphics[width=8.5cm]{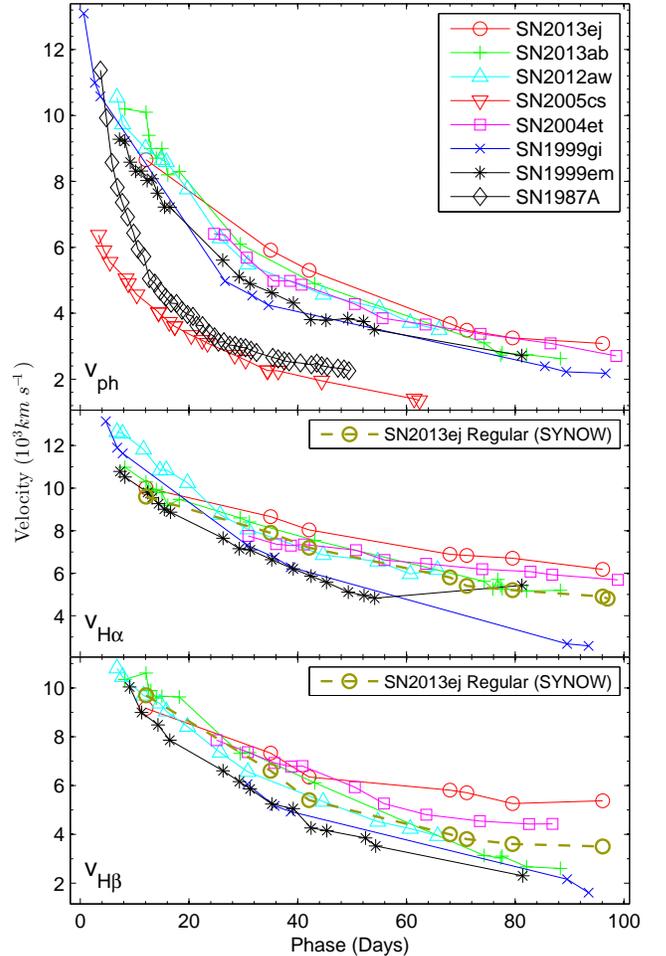}
\caption{The photospheric velocity (top) evolution ($v_{\rm ph}$) of {\sn} is compared with other well-studied type II SNe.
         The $v_{\rm ph}$ plotted here are the absorption trough velocities (average of \Feii\ lines at late
         phases and \Hei\ at early phases). Similar comparison of  P-Cygni absorption velocities, but for \ha\ and \hb\ are shown in middle and bottom panels respectively. The regular velocity component for \ha\ and \hb\ estimated from \synow\ (without HV components; see Table.~\ref{tab:synow}) are also plotted for  comparison.}
\label{fig:sp.velph}
\end{figure}

The comparison of \ha\ and \hb\ velocities with other normal IIP SNe (see Fig.~\ref{fig:sp.velph}), estimated by directly locating the P-Cygni absorption troughs, shows that \sn\ velocities are significantly higher and declines relatively slowly (especially during later phases; 60-100d) as compared to those seen in typical IIP SNe, e.g., 1999em, 1999gi, 2012aw or 2013ab. On the other hand the photospheric velocity comparison with other IIP SNe does not show any such anomaly. This, we suggest as the effect of blending with \Hi\ HV components in \ha\ and \hb, which we could separate out while modeling these broad features with \synow\ having two velocity components. The regular \ha\ and \hb\ velocities estimated from \synow\ declines at a normal rate consistent to that seen in other SNe (see Fig.\ref{fig:sp.velph}), whereas the HV components remains at higher velocities by $ 1000 - 2000 $ \kms, declining at relatively slower rate. It is also interesting to note that the velocity difference between the regular and HV component for \ha\ and \hb\ is similar at same epochs. \cite{2007ApJ...662.1136C} identified similar HV absorption features associated close to \ha\ and \hb\ troughs in SNe 1999em and 2004dj, which remained constant with time. Presence of such HV features has also been detected in SN 2009bw \citep{2012MNRAS.422.1122I} and SN 2012aw \citep{2013MNRAS.433.1871B} which is suggestive of interaction of SN ejecta with pre-existent CSM. Similar to \sn, HV signatures has been detected all throughout the plateau phase evolution of SN 2009bw, while in SN 2012aw such features were only detected at late plateau phase (55 to 104d). Although, we found HV components in \sn\ by modeling the extended P-Cygni troughs, we are unable to visually detect such two individual velocity components, this is possibly because of our signal-to-noise-ratio limited spectra and weaker strength of HV components.
\cite{2007ApJ...662.1136C} argued that SN ejecta can interact with the cooler dense shell of CMS material, which might have originated from the pre-supernova mass loss in the form of stellar winds. Their analysis showed that such interaction can led to the detection of HV absorption features on bluer wings of Balmer lines due to enhanced excitation of the outer layers of unshocked ejecta. We, therefore suggest weak or moderate ejecta-CSM interaction in \sn.
{X-ray emission from \sn\ has also been reported by \cite{2013ATel.5243....1M}, which they measured a 0.3-10 keV count-rate of 2.7$  \pm $0.5 cps, translating into a flux of $ \sim1.1\times10^{-13} $ erg~s$ ^{-1} $cm$ ^{-2} $ (assuming simple power-law spectral model with photon index Gamma $ =2 $). Such X-ray emission may also indicate ejecta-CSM interaction suffered by \sn.}

\section{Status of \sn\ in type II diversity}
\subsection{Factors favoring \sn\ as type IIL}
{Having characterized the event both photometrically and spectroscopically, we may now revisit the aspects which favor \sn\ as type IIL event.
The SN was originally classified as type IIP \citep{2013ATel.5228....1V} based on spectroscopic similarity to SN 1999gi. Due to same underlying physical mechanisms which govern both type IIP and IIL SNe, early spectra may not clearly distinguish these sub classes of SN type II. The distinguishing factor among IIP and IIL is nominal and mainly depend upon light curve characteristics.
\sn\ shows a decline of 1.74 \maghundred\ (see Table~\ref{tab:slopendrop}) or $ \sim0.87 $ mag in 50 days, which definitely falls in the criteria of type IIL SNe as proposed by \cite{2014MNRAS.445..554F}.  In Fig.~\ref{fig:lc.abs}, the spread of template light curves for type IIP and IIL \citep{2014MNRAS.445..554F} is shown along with M$ _V $ light curves of SNe sample. It is evident that under this scheme of classification, \sn\ is not a type IIP, rather it is marginally within the range of type IIL template light curves. This is also justified from the point of basic idea behind these classifications, that type IIP must show a `plateau' of almost constant brightness for some time ($ \sim90$d), which is not the case with \sn.} {Due to the very fact that SN type II light curves and physical properties exhibit a continuum distribution rather than a bi-modality \citep{2014ApJ...786...67A}, \sn\ shows intermediate characteristic in the SN type II diversity.}

{One distinguishing spectroscopic property \cite{2014MNRAS.445..554F} found for type IIL SNe is the overall higher photospheric (\Feii~\ld5196) velocity and flatter \Hi\ (\hb\ and \ha) velocity profiles as compared to type IIP counterpart. Although \Feii\ velocities are on the higher end as compared to typical IIP SNe velocities, we do not find it as a remarkable deviation to distinguish \sn\ from IIP sample. However, we do see a anomaly in \ha, \hb\ absorption minima velocity profiles, as they start off with velocities consistent with those of type IIP but declines relatively slowly (see \S\ref{sec:sp.vel} for more description of this feature) ultimately surpassing faster declining IIP velocity profiles after 50 days. This characteristic feature of \Hi\ velocities for \sn\ is typical for most IIL SNe as found by \cite{2014MNRAS.445..554F}.}

\subsection{CSM interaction and type IIL}
{\cite{2014MNRAS.445..554F} proposed a possible explanation for the flatter velocity profiles in IIL SNe, which is due the lack of hydrogen in deeper and slow expanding layers of ejecta, resulting into higher \Hi\ absorption velocities arising mostly from outer layer. However, for \sn\ we suggest the flattening of \ha\ and \hb\ velocity profiles are due to the contamination of HV component of \Hi\ (see \S\ref{sec:hvcsm}). Indication of CSM interaction in \sn\ may also be inferred from X-ray detection by \cite{2013ATel.5243....1M}.
\cite{2015arXiv150106491V} found SN 2013by, a type IIL SN, to be moderately interacting with CSM. This led them to ask the prevalence of CSM interaction among IIL SNe in general. Type IIL SNe originate from progenitors similar to IIPs, but have lost a significant fraction of hydrogen before explosion during pre SN evolution.
Hence it may not be usual to detect HV \Hi\ signatures in \ha, \hb\ absorption profiles as a consequence of ejecta-CSM interaction. A moderate or weak interaction may produce a HV component blending with \ha, \hb\ profiles, which may result into shift in absorption minima, rather than a prominent secondary HV dip.
Such a scenario may perfectly explain the relatively higher and flatter \Hi\ velocity profiles of most type IIL SNe as compared to IIP counterparts, found by \cite{2014MNRAS.445..554F} based on direct velocity estimates of absorption minima.}

{Another example of CSM interaction in type IIL is SN 2008fq, which does show strong interaction signature like a type IIn \citep{2013A&A...555A..10T}, but also shows a steep decline like IIL during first 60 days \citep{2014MNRAS.445..554F}. {Supernova PTF11iqb \citep{2015MNRAS.449.1876S} is also a type IIn SN, having prominent CSM interaction signatures, but with IIL like steeper light curve. Initial spectra of this SN showed IIn characteristics, however late plateau spectra revealed features similar to type IIL. PTF11iqb originated from a progenitor identical to type IIP/L, instead of a LBV as expected for a typical IIn.} However, because of rare detection of type IIL events and its fast decline in magnitudes we do not have sufficient information to investigate CSM interaction in all such objects. Thus, the question still remains open if all or most IIL SNe interact with CSM and whether the flatter \Hi\ absorption minima velocity profiles is a consequence of interaction.}

\section{Light curve modelling}\label{modelling}
To determine the explosion parameters of \sn, the observed light curve is modeled following the semi-analytical approach originally developed by \cite{1980ApJ...237..541A} and further refined  in \cite{1989ApJ...340..396A}.
More appropriate and accurate approach would have been detailed hydrodynamical modeling \citep[e.g.][]{1977ApJS...33..515F,2007A&A...461..233U,2011ApJ...729...61B,2011ApJ...741...41P} to determine explosion properties, however application of simple semi-analytical models \citep{1980ApJ...237..541A,1982ApJ...253..785A,1989ApJ...340..396A,1993ApJ...414..712P,2003MNRAS.338..711Z,2012ApJ...746..121C} can be useful to get preliminary yet reliable estimates of the parameters without running resource intensive and time consuming hydrodynamical codes. \cite{2014A&A...571A..77N} also followed the original semi-analytical formulation presented by \cite{1989ApJ...340..396A} and modeled a few well studied II SNe. The results are compared with hydrodynamical models from the literature and are found to be in good agreement.
The model light-curve is computed by solving the energy balance of the spherically symmetric supernova envelope, which is assumed to be in homologous expansion having spatially uniform density profile.

The temperature evolution is given as \citep{1980ApJ...237..541A},
\[ T(x,t)^4=T_0^4\psi(x)\phi(t)\left(\frac{R_0}{R(t)}\right)^4 ,\]
where $ x $ is defined as dimensionless co-moving radius relative to the mass of the envelope and,  $ \psi(x) $ is the radial component of temperature profile which falls off with radius as $ sin(\pi x)/\pi x $.
Here we incorporate the effect of recombination, as shock heated and ionized envelope expands and cools down to recombine at temperature $ T_{rec} $. We define $ x_i $ as the co-moving radius of the recombination front and the opacity ($ \kappa $) changes very sharply  at this layer such that $ \kappa \approx 0 $ for the ejecta above $ x_i $. Following the treatment of \cite{1989ApJ...340..396A} the
temporal component of temperature, $ \phi(t) $ can be expressed as \citep{2014A&A...571A..77N},
\[ \frac{d\phi(t)}{dz}= \frac{R(t)}{R_0 x_i^3}\left[p_1\zeta(t)-p_2\phi(t)x_i-2 x_i^2 \phi(t) \frac{R_0}{R(t)}\frac{dx_i}{dz}\right] ,\]
here $ \zeta(t) $ is the total radioactive energy input from decay chain of unit mass of \nickel, which is normalized to the energy production rate of \nickel. The rest of the parameters in the equation have usual meaning and can be found in aforementioned papers. From this ordinary differential equation we can find out the solution of $ \phi(t) $ using Runge-Kutta method. The treatment adopted to determine $ x_i $ is somewhat similar to \cite{2014A&A...571A..77N}, where we numerically determine the radius $ x_i $ (to an accuracy of $ 10^{-12} $) for which the temperature of the layer reaches $ T_{rec} $.  Once we find out the solution of $ \phi(t) $ and $ x_i $, the total bolometric luminosity is calculated as the sum of radioactive heating and rate of energy released due to recombination,
\[ L(t)=x_i\frac{\phi(t)E_{th}(0)}{\tau_d}\left(1-e^{-A_g/t^2}\right)+4\pi r_i^2 Q\rho(x_i,t)R(t)\frac{dx_i}{dt} , \]
here, $ d(x_i)/dt $ is the inward velocity of co-moving recombination front and
the term $ [1-exp(-A_g/t^2)] $, takes into account of gamma-ray leakage from the ejecta.
The factor $ A_g $ is the effectiveness of gamma ray trapping {\citep[see e.g., ][]{1997ApJ...491..375C,2012ApJ...746..121C}}, where large $ A_g $ means full trapping of gamma rays, this factor is particularly important to model the \sn\ tail light curve. In this relation we also modified the second term to correctly account for the amount of envelope mass being recombined.

 \begin{figure}
 \centering
 \hspace{-0.5cm}
 \includegraphics[width=1.05\linewidth]{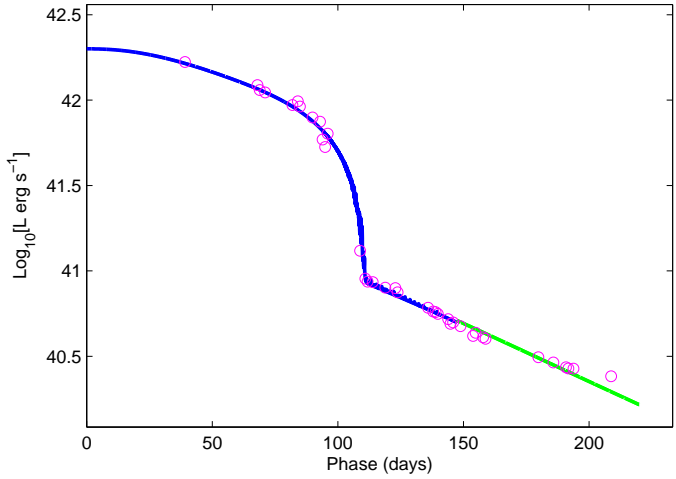}
 \caption{Model fit (solid line) on the observed bolometric light curve (open circles) of \sn. The green solid line follows only the radioactive decay law, where the recombination front has completely disappeared.}
 \label{fig:model}
 \end{figure}

To model SN light curves it is essential to obtain the true bolometric luminosity from observations. Since our data is limited only to optical and UV bands, we adopt the prescription for color dependent bolometric corrections by \cite{2009ApJ...701..200B} to obtain bolometric light curve for \sn. Figure~\ref{fig:model} shows the model fit with the observed bolometric light curve of the SN. We estimate an ejecta mass of 12 \msun, progenitor radius of 450 \rsun\ and explosion energy (kinetic + thermal) of 2.3 foe ($ 10^{51} $ erg). The uncertainty in mass and radius is about 25\%. We find that the plateau duration is strongly correlated with explosion energies (especially kinetic), and also with $ \kappa $ and $ T_{rec} $. Thus depending upon these parameters our model is consistent with a wide range of explosion energies, with 2.3 foe towards the lower end and energies up to 4.5 foe at higher end. Assuming the mass of the compact remnant to be 1.5-2.0 \msun, the total progenitor mass adds up to be 14\msun.

The mass of radioactive \nickel\ estimated from the model is 0.018\msun, which primarily governs the tail light curve of the SN. As discussed in \S\ref{sec:lc.bol}, the slope of the tail light curve observed for \sn\ is significantly higher than other typical IIP SNe and also to that expected from radioactive decay of \cobalt\ to \iron.
The light curve powered by full gamma-ray trapping from radioactive decay chain of $ \nickel \rightarrow \cobalt \rightarrow \iron $ results in a slower decline and does not explain the steeper tail observed in \sn.
In the model we decreased the gamma-ray trapping effectiveness parameter $ A_g $ to $ 3\times10^4 $ day$ ^2 $, which matches the steeper radioactive tail. The gamma-ray optical depth can be related to this parameter as $ \tau_g\sim A_g/t^2 $. This implies that the gamma-ray leakage in \sn\ is significantly higher than other typical type IIP SNe.

\cite{2014MNRAS.438L.101V} using early temperatures ($ <5 $ days) of \sn\ provided a preliminary estimate of the progenitor radius as $ 400-600 $ \rsun, which is in good agreement with our result. Our progenitor mass estimate is also consistent with that reported by \cite{2014MNRAS.439L..56F} from direct observational identification of the progenitor using \textit{HST} archival images, which is $ 8 - 15.5 $ \msun.

\section {Summary} \label{sec:sum}
We present photometric and spectroscopic observations of \sn. Despite low cadence optical photometric follow up during photospheric phase, we are able to cover most of the important phases and features of light curve.

Our high resolution spectrum at 80d shows the presence of \Nai~D (\ldld~5890, 5896) doublet for Milky Way, while no impression for host galaxy \host. This indicates that \sn\ suffers minimal or no reddening due to its host galaxy.

The optical light curves are similar to type IIL SNe, with a relatively short plateau duration of 85d and steeper decline rates of  6.60, 3.57, 1.74, 1.07 and 0.74 mag 100 day$ ^{-1} $ in \textit{UBVRI} bands respectively. The comparison of absolute \textit{V} band light curves shows that \sn\ suffers the higher decline rate than all type IIP SNe, but similar to type IIL SNe 1980k, 2000dc and 2013by.
The drop in luminosity during the plateau-nebular transition is also higher than most type II SNe in our sample, which is 2.4 mag in \textit{V} band.

The UVOT UV optical light curves shows steep decline during first 30 days at a rate of 0.182, 0.213, 0.262  mag d$ ^{-1} $ in \textit{uvw1, uvw2} and \textit{uvm2} bands respectively. The absolute UV light curves are identical to SN 2012aw and also shows a similar UV-plateau trend as observed in SN 2012aw.

Owing to the large drop in luminosity during plateau-nebular transition, the light curve settles to a significantly low luminous tail phase as compared to other normal IIP SNe. The mass of radioactive \nickel\ estimated from the tail bolometric luminosity is $ 0.020\pm0.002 $ \msun, which is in between normal IIP SNe (e.g., 1999em, 2004et, 2012aw) and subluminous events, like SN 2005cs.

The spectroscopic features and their evolution is similar to normal type II events. Detailed \synow\ modelling has been performed to identify spectral features and to estimate velocities for \ha, \hb, \Feii\ (\ldld~4924, 5018, 5169) and \Scii\ (\ldld~4670, 6247) lines. The photospheric velocity profile of \sn, which is represented by \Feii\ lines and \Hei\ line at 12d, is almost identical to SNe 2004et, 2012aw and 2013ab.
The \ha, \hb\ velocities estimated by directly locating the absorption troughs are significantly higher and slow declining as compared to other normal IIP events. However, such \Hi\ velocity profiles are typical for type IIL SNe.

{The P-Cygni absorption troughs of \ha\ and \hb\ are found to be broad and extended which a single \Hi\ component in \synow\ model could not fit properly. However, these extended features are fitted well with \synow\ by incorporating a high velocity \Hi\ component.
These HV components can be traced throughout the photospheric phase which may indicate possible ejecta-CSM interaction.
Our inference is also supported by the detection of X-ray emission from the \sn\ \citep{2013ATel.5243....1M} indicating possible CSM interaction, and the unusually high polarization reported by \cite{2013ATel.5275....1L} may also further indicate asymmetry in environment or ejecta of the SN.
Such CSM interaction and their signature in \ha, \hb\ profiles has also been reported for SNe 2009bw \citep{2012MNRAS.422.1122I} and 2012aw \citep{2013MNRAS.433.1871B}.}

Nebular phase spectra during 109 to 125d phases are dominated by characteristic emission lines, however the \ha\ line shows an unusual notch, which may be explained by superposition of HV emission on regular \ha\ profile. Although, the origin of the feature is not fully explained, it may indicate bipolar distribution of \nickel\ in the core.

We modeled the bolometric light curve of \sn\ and estimated a progenitor mass of $ \sim14 $\msun, radius of $ \sim450 $\rsun\ and explosion energy of $ \sim2.3$ foe. These progenitor property estimates are consistent to those given by \cite{2014MNRAS.439L..56F} and \cite{2014MNRAS.438L.101V} for mass and radius respectively. The tail bolometric light curve of \sn, is found to be significantly steeper than that expected from decay chain of radioactive \nickel. Thus, in the model we decreased the effectiveness of gamma ray trapping, which could explain the steeper slope of tail light curve.

\acknowledgments
We are thankful to the observing staffs and technical assistants of ARIES 1.0-m and 1.3-m telescopes and we also express our thanks to 2-m HCT telescope staffs for their kind cooperation in observation of \sn. We also express our thanks to Mr. Shashank Shekhar for his sincere efforts and co-operation during observations at ARIES 1.3m telescope. Authors gratefully acknowledge the services of the
NASA ADS and NED databases which are used to access data and references in this paper. SOUSA is supported by NASA's Astrophysics Data Analysis Program through grant NNX13AF35G. VVDs work on Type IIP SNe is supported by the NASA through Chandra Award Number GO2-13092B issued by the Chandra X-ray Observatory Center, which is operated by the Smithsonian Astrophysical Observatory for and on behalf of the NASA under contract NAS8-03060.
We also thank the anonymous referee for detailed and insightful comments which helped in significant improvement of the manuscript.

\bibliography{ms}

\clearpage
\appendix
\section{Nebular \ha\ profile}
\label{app:nebular_ha}
\begin{figure}
\centering
\includegraphics[width=0.35\linewidth]{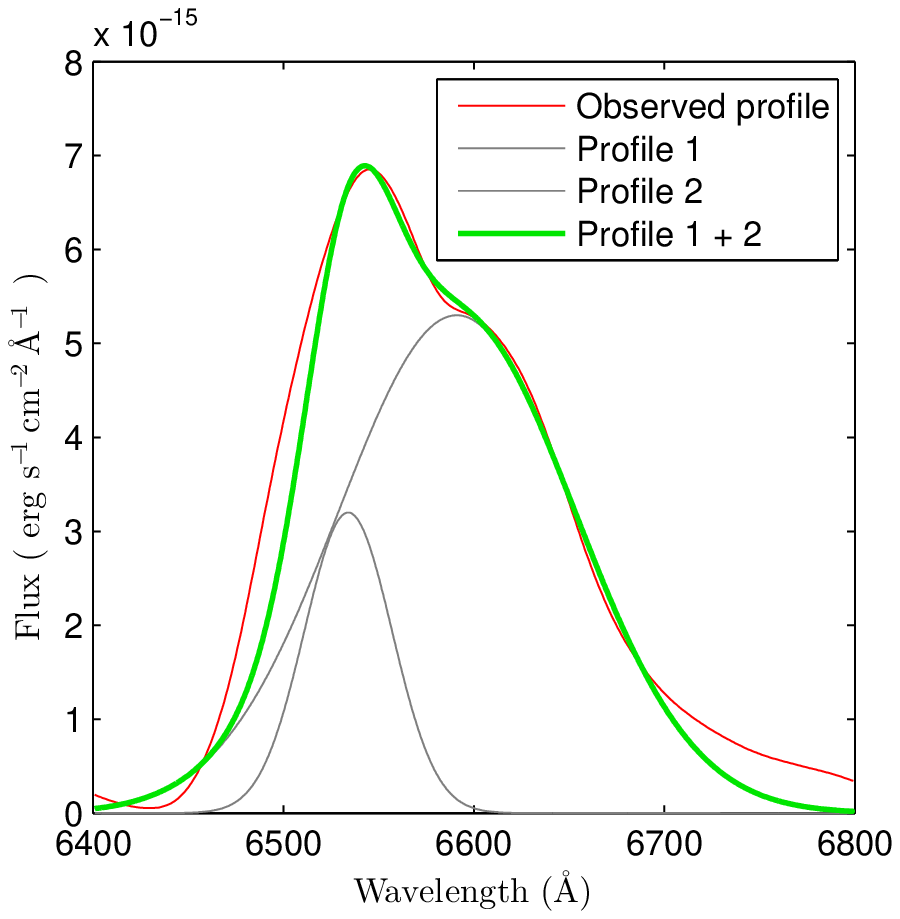}
\caption{\ha\ profile of 125d spectrum fitted by two component gausian profile seperated by $ \sim $2500 \kms.}
\label{fig:profile_combi}
\end{figure}

{The unusual notch in nebular \ha\ profile can be described as a superimposition of two profiles. Fig.~\ref{fig:profile_combi} shows the observed \ha\ profile at 125d which is fitted by two component Gaussian profiles. These two profiles are separated by 55 \AA\ ($ \sim2500 $ \kms), one being blue shifted by -1300 \kms\ while the other is red shifted at 1200 \kms\ with respect to rest \ha\ position. The FWHM for the blue component is 54\AA\ and for red component is 146\AA. The redshifted component is dominant in strength over the blue one, having their ratio of equivalent widths to be 4.5.}
It may be noted that for the sake of simplicity and only for the purpose of illustration we used Gaussian profiles, which does not account for the P-Cygni absorption troughs as we see on bluer wings of line profiles in observed SN spectrum.

\end{document}